\documentclass{appolb}
\usepackage{amssymb,amsmath,bm}

\usepackage{epsf}
\usepackage{epsfig,rotate}
\usepackage{afterpage}
\usepackage{longtable}
\usepackage{cite}
\usepackage{color}

\newcommand{\vus}{|V_{us}|}
\newcommand{\vcb}{|V_{cb}|}
\newcommand{\vtd}{|V_{td}|}
\newcommand{\vub}{|V_{ub}|}
\newcommand{\vts}{|V_{ts}|}

\def\epe{\varepsilon'/\varepsilon}
\newcommand{\tev}{\, {\rm TeV}}
\newcommand{\gev}{\, {\rm GeV}}
\newcommand{\mev}{\, {\rm MeV}}

\newcommand{\ts}{\tilde s}
\newcommand{\tc}{\tilde c}
\newcommand{\Vt}{\widetilde V}

\newcommand{\be}{\begin{equation}}
\newcommand{\ee}{\end{equation}}
\newcommand{\bea}{\begin{eqnarray}}
\newcommand{\eea}{\end{eqnarray}}

\newcommand{\ba}{\begin{array}}
\newcommand{\ea}{\end{array}}

\newcommand{\bbs}{\ensuremath{B_s\!-\!\overline{\!B}_s\,}}
\newcommand{\bbms}{\bbs\ mixing}
\newcommand{\bbd}{\ensuremath{B_d\!-\!\overline{\!B}_d\,}}
\newcommand{\bbmd}{\bbd\ mixing}
\newcommand{\kk}{\ensuremath{K\!-\!\overline{\!K}\,}}
\newcommand{\kkm}{\kk\ mixing}

\newcommand{\ord}{{\cal O}}

\def\kpn{K^+\rightarrow\pi^+\nu\bar\nu}

\def\klpn{K_{L}\rightarrow\pi^0\nu\bar\nu}

\begin{document}
\headtitle{BSM Models Facing the Recent LHCb Data 
\dots}
\headauthor{Andrzej J.~Buras and Jennifer Girrbach}
\title{
  BSM MODELS FACING THE RECENT LHCb DATA: \\ A FIRST LOOK
\thanks{Updated talk presented by A.J.B. at the  Cracow Epiphany Conference
``On Present and Future of B-Physics'', Cracow, Poland, January 8-11, 2012.}
}
\author{Andrzej J. Buras and Jennifer Girrbach
\address{Technical University Munich, Physics Department, D-85748 Garching, Germany,\\
TUM-IAS, Lichtenbergstr. 2a, D-85748 Garching, Germany
 \\
   }}
\maketitle
\begin{abstract}
During last decade a number of detailed analyses of flavour 
observables and of their correlations within more than a dozen specific BSM models 
have been performed at the TUM. One of the goals of these analyses was 
to investigate which model is capable of obtaining large mixing induced 
CP asymmetry in the $B_s$ system, $S_{\psi\phi}$, and to find out what this 
would imply for other flavour observables. In this context also the rare decays $B_{s,d}\to\mu^+\mu^-$ have been considered. In some 
models their branching ratios can be enhanced by  
 orders of magnitude above the SM expectations. The recent data 
on  $S_{\psi\phi}$ and  $B_{s,d}\to\mu^+\mu^-$ from the LHCb put an end 
to these very optimistic hopes modifying significantly the allowed 
patterns of deviations from SM predictions for flavour observables in 
concrete BSM models. We make a first semi-quantitative 
assessment of the most important modifications in the predictions of the BSM 
models in question including also recently analyzed models and taking 
into account the most recent lattice input. Our presentation is 
dominated by quark flavour observables in $B_{s,d}$ and $K^+(K^0)$ meson 
systems.
For some BSM models the LHCb data turned out to be a relief. 
On the other hand the SM,
models with CMFV and  MFV models without flavour blind phases 
appear to have significant difficulties in describing all $\Delta F=2$ 
observables in $B_{s,d}$ and $K^0$ meson 
systems simultaneously. However, definite conclusions will only 
be possible when $\vub$ and $\gamma$ will be known from tree level decays 
with a much better precision and the lattice input will  further improve.
 Finally, we propose to regard the stringent CMFV relations between 
various observables as {\it 
standard candels of flavour physics.}  The pattern of deviations from these relations  may help 
in identifying the correct NP scenario.\\
\vspace{0.5cm}\\
 FLAVOUR(267104)-ERC-10
\end{abstract}

\newpage

\section{Introduction}

 This decade should provide a much better understanding of the physics 
at the shortest distance scales explored by humans, that is scales of order 
$5\times 10^{-20}$m   explored by ATLAS and CMS and possibly even shorter distance
scales explored by dedicated flavour physics experiments like LHCb, 
SuperKEKB, SuperB in Rome and Kaon physics dedicated experiments like NA62, 
${\rm K^0}$TO and ORCA. The main goal of these experiments is the search for 
New Physics (NP).

The most efficient way to uncover NP in weak decay processes 
is to 
identify correlations between flavour observables characteristic 
for a given extension of the SM. Such correlations being less 
sensitive to the model parameters can often allow a transparent 
distinction between various models proposed in the literature 
\cite{Buras:2010wr}. Also the so-called ``DNA-Test'' of a given 
model can give a global insight into the particular pattern of 
possible deviations from SM predictions for flavour observables 
\cite{Altmannshofer:2009ne}. Such studies are at the frontiers 
in our search for a fundamental theory of elementary particles 
in which flavour violating interactions will play undoubtedly a prominent 
role
\cite{Buras:2009if,Isidori:2010kg,Fleischer:2010qb,Nir:2010jr,Antonelli:2009ws,Hurth:2010tk,Ciuchini:2011ca,Meadows:2011bk}. 

Now until recently the number of flavour observables used efficiently to test 
the short distance structure of the SM and of various BSM scenarios was 
limited to tree level decays, particle-antiparticle mixing including 
CP-violating observables like $\varepsilon_K$ and $S_{\psi K_S}$ and a handful 
of $\Delta F=1$ loop induced processes like $B\to X_s\gamma$ and 
$B\to X_s\ell^+\ell^-$.  This allowed in many models still
significant deviations from the  SM expectations  summarized 
in \cite{Buras:2010wr}. In particular a number of
models, having a multitude of free parameters were capable of 
obtaining large mixing induced 
CP asymmetry in the $B_s$ system, $S_{\psi\phi}$, signaled initially 
by CDF and D0.  This in turn implied  often significant NP effects 
in some flavour observables and/or precluded in certain models large NP effects 
in other observables, in particular in rare Kaon decays like 
$\kpn$ and $\klpn$.
In these analyses the rare decays $B_{s,d}\to\mu^+\mu^-$ have also
played a prominent role. In fact
in models with new heavy neutral scalars like MSSM and 2HDMs of various type
their branching ratios could  in the spring of 2011 still be enhanced by one  
 order of magnitude above the SM expectations while satisfying all 
existing data. 

While already the messages from Tevatron last  summer indicated that NP effects 
are probably smaller in $B_s$ flavour physics than initially expected and 
hoped for,
the very recent data 
on  $S_{\psi\phi}$ and  $B_{s,d}\to\mu^+\mu^-$ from the LHCb put an end 
to these very optimistic hopes modifying significantly the allowed 
patterns of deviations from SM predictions for flavour observables in 
concrete BSM models. 

Indeed, the most recent data on $S_{\psi\phi}$ and
$B_{s,d}\to\mu^+\mu^-$ decays from LHCb read \cite{Aaij:2012ac}
\be\label{LHCb1}
S_{\psi\phi}=0.002\pm 0.087, \quad S^{\rm SM}_{\psi\phi}=0.035\pm 0.002,
\ee
\be\label{LHCb2}
\mathcal{B}(B_{s}\to\mu^+\mu^-) \le 4.5\times 10^{-9}, \quad
\mathcal{B}(B_{s}\to\mu^+\mu^-)^{\rm SM}=(3.1\pm0.2)\times 10^{-9},
\ee
\be\label{LHCb3}
\mathcal{B}(B_{d}\to\mu^+\mu^-) \le 8.1\times 10^{-10}, \quad
\mathcal{B}(B_{d}\to\mu^+\mu^-)^{\rm SM}=(1.0\pm0.1)\times 10^{-10},
\ee
where we have also shown updated SM predictions  for these 
observables as discussed in Section~\ref{sec:3}. They differ only marginally from 
those quoted in \cite{Buras:2010wr} and given in (\ref{BRSM2}) and 
(\ref{BRSM1}). Our phase sign convention is such that in the 
SM  $S_{\psi\phi}$ is positive. See also (\ref{eq:3.43}).
The experimental 
error on $S_{\psi\phi}$ has been obtained by adding statistical and systematic 
errors in quadrature and the upper bounds on $\mathcal{B}(B_{s,d}\to\mu^+\mu^-)$  are at $95\%$ C.L.

Indeed it looks like the SM still survived another test: 
mixing induced CP-violation in $B_s$ decays is significantly smaller than in 
$B_d$ decays as expected in the SM already for 25 years. However from the present perspective $S_{\psi\phi}$ could still be found
in the range 
\be\label{Spsiphirange}
-0.20\le S_{\psi\phi}\le 0.20
\ee
and finding it to be negative would be not only a clear signal of NP but 
would rule out a number of models as we will see below. Moreover finding 
it above $0.1$ would also be a signal of NP but not as pronounced as a 
negative value. 

Concerning $B_{s}\to\mu^+\mu^-$, as pointed out recently in \cite{deBruyn:2012wj,deBruyn:2012wk}, when 
comparing the theoretical branching ratio in (\ref{LHCb2}) with experiment, 
a correction factor has to be included which takes care of $\Delta\Gamma_s$ 
effects that influence the extraction of this branching ratio from the data:
\be\label{Fleischer}
\mathcal{B}(B_{s}\to\mu^+\mu^-)_{\rm th}=r(\Delta\Gamma_s)\mathcal{B}(B_{s}\to\mu^+\mu^-)_{\rm exp}, \quad r(0)=1.
\ee
The authors of \cite{deBruyn:2012wj,deBruyn:2012wk}
find $r(\Delta\Gamma_s)=0.91\pm0.01$. It is a matter of 
choice whether this factor should be included in the theoretical calculation 
or in the experimental branching ratio. We prefer to include it in the latter 
so that the experimental upper bound in (\ref{LHCb2}) is reduced by $9\%$ implying a more stringent 
upper bound of $4.1\times 10^{-9}$. Thus finally the SM central value is only by a 
factor of $1.3$ below the experimental $95\%$ C.L. upper bound.

Unfortunately this  bound on
$\mathcal{B}(B_{s}\to\mu^+\mu^-)$ precludes a simple distinction between NP 
contributions coming from neutral scalars and neutral gauge bosons which 
would be possible if its value was larger than $6\cdot 10^{-9}$.  Indeed such 
an enhancement  
could most easily be attributed to heavy neutral scalar exchanges
\cite{Altmannshofer:2011gn}.
On the other hand 
we should note that the upper bound on 
$\mathcal{B}(B_{d}\to\mu^+\mu^-)$ is still one order of magnitude above its
SM value and due to the smallness of $\Delta\Gamma_d$ 
is not affected by the correction discussed above. It could turn out after all that it is $B_{d}\to\mu^+\mu^-$ 
and not  $B_{s}\to\mu^+\mu^-$ that will most clearly signal NP in these 
decays. We will return to this point below.

However, the recent LHCb data listed above are not the only flavour 
highlights of the last half a year that are relevant for our presentation. 
In particular improved values on the five non-perturbative parameters 
\cite{Laiho:2009eu}\footnote{We thank Christine Davies for help in getting a better insight  into these results.}
\be\label{lattice1}
\hat B_K= 0.767(10), \quad F_{B_d}=(190.6\pm4.6)\mev, \quad F_{B_s}=(227.7\pm6.2)\mev,  
\ee
and
\be\label{lattice2}
\xi=\frac{F_{B_s} \sqrt{\hat B_{B_s}}}{F_{B_d} \sqrt{\hat B_{B_d}}}= 
1.237\pm0.032, \quad
F_{B_s} \sqrt{\hat B_{B_s}}=279(13)\mev
\ee
allowed for improved SM predictions
for $\varepsilon_K$, $\Delta M_{s,d}$ and  $\mathcal{B}(B_{s,d}\to\mu^+\mu^-)$.
The values quoted above are taken from a recent update of lattice 
averages in \cite{Laiho:2009eu} that are based on a number of 
lattice calculations for which the references can be found in this 
paper. In particular for $B_{s,d}-\bar B_{s,d}$ physics one should refer to an impressive 
precision reached in \cite{Gamiz:2009ku,Na:2012kp,Davies:2012qf,Bouchard:2011xj} and 
in the case of $K^0-\bar K^0$ to \cite{Antonio:2007pb}\footnote{The first author of 
this review expects on the basis of \cite{Gerard:2010jt,Bardeen:1987vg} that the value of 
$\hat B_K$ will eventually be below $0.75$ or equal to it.}

In (\ref{lattice2}) 
following the recommendations of lattice experts we use $\xi$ and 
$F_{B_s} \sqrt{\hat B_{B_s}}$ as basic lattice input which gives
\be
F_{B_d} \sqrt{\hat B_{B_d}}= 226(13)\mev, 
\ee
that agrees well with the direct average in \cite{Laiho:2009eu}
$F_{B_d} \sqrt{\hat B_{B_d}}= 227(17)\mev$.

The goal of this writing is to summarize first the possible deviations from 
SM predictions in flavour observables and investigate which models can 
remove these anomalies while being consistent with all available data, 
in particular with the LHCb data quoted above. However as already advertised 
in the title, our main goal is to confront the models reviewed in
\cite{Buras:2010wr} and few other models studied since then, with the 
LHCb data above and to investigate how the most important predictions of 
these models are modified by these new messages from the nature. 
A recent analysis of the implications of the LHCb data on NP in
$B^0_{d,s}-\bar B^0_{d,s}$ mixings 
 has been 
presented   from a different perspective in \cite{Lenz:2012az}.

In view of space limitations we have to develop a strategy for presenting 
our observations and results. First,  as far as CP-violating observables are concerned the main stars of our presentation will be:
\be\label{MAINSTARSCP}
\varepsilon_K, \quad S_{\psi K_S}, \quad  S_{\psi\phi}, \quad \mathcal{B}(\klpn).
\ee
Among CP-conserving ones
we will pay particular attention to
\be\label{MAINSTARSCPC}
\Delta M_{s,d}, \quad \mathcal{B}(B_{s,d}\to \mu^+\mu^-),\quad 
\mathcal{B}(B^+\to \tau^+\nu_\tau), \quad
\mathcal{B}(\kpn). 
\ee
This means that several important decays like $B\to K^*\ell^+\ell^-$ and 
all $b\to s \nu\bar\nu$ transitions will barely appear on the scene. We 
hope to improve on this in the future. For the time being we refer 
to   \cite{Altmannshofer:2011gn} and references therein. Similarly we 
will not discuss CP-violation in charm decays as in view of recent 
LHCb data charm experts have already presented several views in the 
literature and we have nothing to add here at present.
 Concerning lepton flavour 
violation it will only show up in GUT models.

Our discussion will be at best semi-quantitative as a full-fledged analysis 
would require redoing all numerical analyses reviewed in \cite{Buras:2010wr} 
putting new constraints on the masses of new particles in these models that 
are being obtained from the LHC. This is clearly beyond the scope of this 
rather
short review. Moreover, while the lower bounds on the masses of new 
particles increased during last year, in evaluating branching ratios 
this increase can be approximately compensated by the increase of the 
relevant couplings and mixing parameters so that the picture obtained 
through our rough analysis should be roughly correct. The reason why 
we  can at all make any statements about the changes in predictions of 
various models without basically performing any new numerical 
analysis is that 
the strategy of the analyses discussed in \cite{Buras:2010wr} was to 
present the predictions for various observables as functions of $S_{\psi\phi}$. 
Therefore simply inspecting the plots from different analyses in the corresponding 
papers one can get a rough 
idea of what is going on after new LHCb results have been taken into account. 
Other correlations presented there, involving this time the branching ratios 
$\mathcal{B}(B_{s,d}\to \mu^+\mu^-)$, turned out to be very helpful 
in this respect as well.

Finally, our strategy won't be to repeat  any details of the models considered here, because 
 in \cite{Buras:2010wr} a rather compact presentation of most of them
 can be found. Interested readers are asked to read in parallel  \cite{Buras:2010wr} and 
related original papers. 

Our review has a very simple structure. In Section~\ref{sec:2} we 
summarize briefly the present anomalies in the flavour data as seen from 
the point of view of the SM. In Section~\ref{sec:3}, the main section of 
this writing, we will have a first look at the modifications of the results 
of all models considered in \cite{Buras:2010wr}, adding few models that 
have been analysed by us since then. A brief outlook consisting of observations, messages and 
a shopping list for coming years  in Section~\ref{sec:4} 
ends our review.

\section{Anomalies in the Flavour Data}\label{sec:2}
Let us summarize 
the pattern of deviations from the SM expectations presently observed in 
the data. In this context it should be emphasized that because of the 
$\varepsilon_K-S_{\psi K_S}$ tension \cite{Lunghi:2008aa,Buras:2008nn,Buras:2009pj,Lunghi:2009sm,Lunghi:2010gv,UTfit-web,Laiho:2012ss} within
the SM this pattern depends on 
whether $\varepsilon_K$ or $S_{\psi K_S}$ is used as a basic observable to 
fit the CKM parameters. As both observables can receive important 
contributions from NP, none of them is optimal for this goal. The solution 
to this problem will be  solved one day by precise measurements of the CKM parameters 
with the help 
of tree-level decays. Unfortunately, the tension between the inclusive 
and exclusive determinations of $\vub$ and the poor knowledge of  the 
phase $\gamma$ 
from tree-level decays preclude this solution at present. In view of this 
it is useful to set $\gamma\approx 70^\circ$ and consider two scenarios for $\vub$:
 \begin{itemize}
\item
{\bf Exclusive (small) $\vub$ Scenario 1:}
$|\varepsilon_K|$ is smaller than its experimental determination,
while $S_{\psi K_S}$ is very close to the central experimental value.
\item
{\bf Inclusive (large) $\vub$ Scenario 2:}
$|\varepsilon_K|$ is consistent with its experimental determination,
while $S_{\psi K_S}$ is significantly higher than its  experimental value.
\end{itemize}

Thus dependently which scenario is considered we need either  
{\it constructive} NP contributions to $|\varepsilon_K|$ (Scenario 1) 
or {\it destructive} NP contributions to  $S_{\psi K_S}$ (Scenario 2). 
However this  NP should not spoil the agreement with the data 
for $S_{\psi K_S}$ (Scenario 1) and for $|\varepsilon_K|$ (Scenario 2).

In view of the fact that the theoretical precision on $S_{\psi K_S}$ 
is significantly larger than in the case of $\varepsilon_K$, one may wonder 
whether removing $1-2\sigma$ anomaly in $\varepsilon_K$ 
by generating a $2-3\sigma$ anomaly in $S_{\psi K_S}$ is a reasonable 
strategy. However, one should take into account that in addition to 
the fact that large values of $\vub$ are found in inclusive $B$-decays 
there is still another tension within the SM that similarly to $\varepsilon_K$
favours large $\vub$ scenario:
\begin{itemize}
\item
The SM branching ratio for $B^+\to\tau^+\nu_\tau$ in Scenario 1 is 
by a factor of two below the data, although the latter are not very 
precise and one can talk only about a $2.5\sigma$ discrepancy. In Scenario 2 
the discrepancy is much smaller, about $1\sigma$. Models 
providing an {\it enhancement} of this branching ratio should be definitely 
favoured in the case of Scenario 1 but in Scenario 2 this is not so 
evident in view of large experimental error.
\end{itemize}
In any case we think it is useful to concentrate on these two NP scenarios, 
even if precise definition of these scenarios depends on particular value 
of $\vub$. We will be more specific about it in our numerical examples 
below.

Now models with many new parameters can face successfully both scenarios 
removing the deviations from the data for certain ranges of their parameters 
but as we will see below in simpler models often 
only one scenario can be admitted as only in that scenario for $\vub$ a given 
model has a chance to fit $\varepsilon_K$ and $S_{\psi K_S}$ simultaneously. 
For instance as we will see in the next section models with constrained 
MFV select Scenario 1, while the 2HDM with MFV and flavour blind phases,
${\rm 2HDM_{\overline{MFV}}}$, favours  Scenario 2 for $\vub$.
What is interesting is that the future precise determination of $\vub$ through 
tree level decays will 
be able to distinguish between these two NP scenarios. We will see that there 
are other models which can be distinguished in this simple manner.

Now the tensions within the SM discussed above constitute only a subset 
of visible deviations of its predictions from the data. A 
closer look at the measured quark flavour observables reveals the following deviations 
at the  $1-2~\sigma$ level:
\begin{itemize}
\item
The  SM branching ratio for the inclusive decay $B\to X_s\gamma$ is by 
$1.2\sigma$ below the data so that models providing an  {\it enhancement} 
appear to be favoured.
\item
The SM inclusive branching ratio for $B\to X_s\ell^+\ell^-$ at high $q^2$ is 
visibly below the data.
\item
The $K^*$ longitudinal polarisation fraction $F_L$ in $B\to K^*\ell^+\ell^-$ predicted by the SM is 
on the other hand larger than the data.
\item
The asymmetry $A_{\rm SL}^b$ measured by D0 is by $3.9\sigma$ different from 
the SM value.
\end{itemize}

In Table~\ref{tab:SMpred} we illustrate the SM predictions for some of 
these observables in both scenarios setting $\gamma=68^\circ$.  What is 
striking in this table is that with the new lattice input in (\ref{lattice2})
the predicted central values of $\Delta M_s$  and $\Delta M_d$, although 
slightly above the data,  are both in a good agreement with the latter 
when hadronic uncertainties are taken into account. In particular 
the central value of the ratio $\Delta M_s/\Delta M_d$ is 
 very close to  the data. 
These results depend strongly on the lattice input and in the case 
of $\Delta M_d$ on the value of $\gamma$. Therefore to get a better insight 
both lattice input and the tree level determination of $\gamma$ 
have to improve. 
From 
the present perspective, models providing  $10\%$ {\it suppression} 
of {\it both} $\Delta M_s$ and $\Delta M_d$ 
appear to be slightly favoured.

\begin{table}[!tbh]
\centering
\begin{tabular}{|c||c|c|c|}
\hline
 & Scenario 1: & Scenario 2:   & Experiment\\
\hline
\hline
  \parbox[0pt][1.6em][c]{0cm}{} $|\varepsilon_K|$ & $1.87(26)  \cdot 10^{-3}$  & $2.28(32)\cdot 10^{-3}$ &$ 2.228(11)\times 10^{-3}$ \\

 \parbox[0pt][1.6em][c]{0cm}{}$\mathcal{B}(B^+\to \tau^+\nu_\tau)$&  $0.74(14) \cdot 10^{-4}$&$1.19(20)\cdot 10^{-4}$ & $1.73(35) \times
10^{-4}$\\
 \parbox[0pt][1.6em][c]{0cm}{}$(\sin2\beta)_\text{true}$ & 0.676(25) &0.812(23)  & $0.679(20)$\\
 \parbox[0pt][1.6em][c]{0cm}{}$\Delta M_s\, [\text{ps}^{-1}]$ &19.0(21)&  19.1(21) &$17.77(12)$ \\
 \parbox[0pt][1.6em][c]{0cm}{} $\Delta M_d\, [\text{ps}^{-1}]$ &0.55(6) &0.56(6)   &  $0.507(4)$\\
 \parbox[0pt][1.6em][c]{0cm}{}$\mathcal{B}(B\to X_s\gamma)$ &$3.15(23) \cdot 10^{-4}$&$3.15(23)\cdot 10^{-4}$ & $3.55(26) \times
10^{-4}$\\
\hline
\end{tabular}
\caption{\it SM prediction for various observables for  $|V_{ub}|=3.4\cdot 10^{-3}$ and $|V_{ub}|=4.3\cdot 10^{-3}$ and $\gamma =
68^\circ$ compared to experiment. 
}\label{tab:SMpred}
\end{table}

We are aware of the fact that all these deviations are not yet very significant 
and could disappear. However, for the purpose of our presentation, it 
is useful to take them first seriously keeping in mind that the pattern 
of deviations from SM expectations could be modified in the future. 
This is in particular the case of observables, like $\Delta M_{s,d}$, that 
still suffer from non-perturbative uncertainties. 
It could turn out that suppressions (enhancements) of some observables 
 required presently from NP 
will be modified to enhancements (suppressions) in the future.

\begin{figure}[!tb]
\centerline{\includegraphics[width=0.65\textwidth]{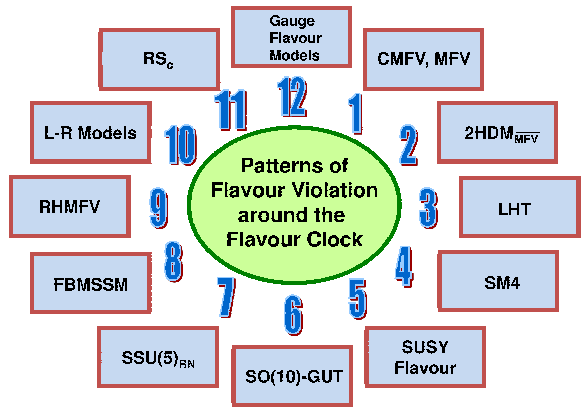}}
\caption{Various patterns of flavour violation around the Flavour Clock.}\label{Fig:2}
\end{figure}

\section{A New Look at BSM Scenarios}\label{sec:3}
\subsection{Preliminaries}
In this section we will make a new look at a number of models analyzed 
in detail in the last decade in Munich in order to see how their patterns 
of flavour violation and CP-violation are affected by the recent LHCb data.
We will first of all discuss models that have been reviewed in some detail in 
 \cite{Buras:2010wr}.
These are:
 CMFV, MFV, ${\rm 2HDM_{\overline{MFV}}}$, 
the Littlest Higgs model with T-parity (LHT), the SM with sequential fourth 
generation (SM4), four classes of supersymmetric flavour models (SF),
supersymmetric SU(5) GUT enriched through RH neutrinos $SSU(5)_{\rm RN}$,
flavour blind MSSM (FBMSSM), the minimal effective model with right-handed 
currents (RHMFV) and
Randall-Sundrum model with custodial protection (RSc). In 2011 new models 
could be added to this list: left-right symmetric model based on 
the electroweak gauge group $SU(2)_L\times SU(2)_R\times U(1)_{B-L}$, 
a maximal gauged flavour model (MGF) and a $SO(10)$-GUT.

Some of these analyses included also lepton flavour violations, EDMs and 
 $(g-2)_\mu$ but in the presentation below we will concentrate  dominantly
on flavour violating and CP-violating processes in the quark 
sector. 
The models to be discussed below are summarized  in order of presentation 
in Fig.~\ref{Fig:2}.
\subsection{CKM Parameters}
In our numerical examples we will use 
\be
\vus=\lambda=0.2252, \qquad \vcb=0.0406, 
\ee
which have been determined by means of 
tree level decays. The values of $\vub$ and $\gamma$ will be specified 
in the context of our presentation.

We recall that once these four parameters of the CKM matrix have been fixed, the {``true''} values 
of the angle $\beta$  and 
of the element $\vtd$ 
are  obtained from the unitarity of the CKM matrix:
\begin{equation} \label{eq:Rt_beta}
\vtd=\vus \vcb R_t,\quad
R_t=\sqrt{1+R_b^2-2 R_b\cos\gamma} ~,\quad
\cot\beta=\frac{1-R_b\cos\gamma}{R_b\sin\gamma}~,
\end{equation}
where
\be
 R_b=\left(1-\frac{\lambda^2}{2}\right)\frac{1}{\lambda}\frac{|V_{ub}|}{\vcb}.
\ee

\subsection{Constrained Minimal Flavour Violation (CMFV)}
The simplest class of extensions of the SM are  models with 
CMFV
\cite{Buras:2000dm, Buras:2003jf,Blanke:2006ig}.
They are formulated as follows:
\begin{itemize}
\item
All flavour changing transitions are governed by the CKM matrix with the 
CKM phase being the only source of CP violation,
\item
The only relevant operators in the effective Hamiltonian below the weak scale
are those that are also relevant in the SM.
\end{itemize}

There are basically three main implications of these assumptions:
\begin{itemize}
\item
$S_{\psi K_S}$ and $S_{\psi\phi}$ are as in the SM and therefore given 
by 
\be
S_{\psi K_S} = \sin(2\beta)\,,\qquad S_{\psi\phi} =  \sin(2|\beta_s|)\,,
\label{eq:3.43}
\ee
where $\beta$ and $\beta_s$ are defined by
\be\label{vtd}
V_{td}=\vtd e^{-i\beta}, \qquad V_{ts}=-\vts e^{-i\beta_s}.
\ee
\item 
For fixed CKM parameters determined in tree-level decays, $|\varepsilon_K|$, 
$\Delta M_s$ and $\Delta M_d$, if modified,  can only be {\it enhanced} 
relative to SM predictions  \cite{Blanke:2006yh}. Moreover this 
happens in a correlated manner \cite{Buras:2000xq}.
\item 
There are relations between various observables that are valid for the  full
class of the CMFV models including the SM. A review of these relations is 
given in \cite{Buras:2003jf}.
We will list some of them now.
\end{itemize}

The most interesting relations in question are the following ones:
\be\label{dmsdmd}
\frac{\Delta M_d}{\Delta M_s}=
\frac{m_{B_d}}{m_{B_s}}
\frac{\hat B_{d}}{\hat B_{s}}\frac{F^2_{B_d}}{F^2_{B_s}}
\left|\frac{V_{td}}{V_{ts}}\right|^2r(\Delta M)
=\frac{m_{B_d}}{m_{B_s}}
\frac{1}{\xi^2}
\left|\frac{V_{td}}{V_{ts}}\right|^2r(\Delta M)
\end{equation}
\begin{equation}\label{bxnn}
\frac{\mathcal{B}(B\to X_d\nu\bar\nu)}{\mathcal{B}(B\to X_s\nu\bar\nu)}=
\left|\frac{V_{td}}{V_{ts}}\right|^2r(\nu\bar\nu)
\end{equation}
\begin{equation}\label{bmumu}
\frac{\mathcal{B}(B_d\to\mu^+\mu^-)}{\mathcal{B}(B_s\to\mu^+\mu^-)}=
\frac{\tau({B_d})}{\tau({B_s})}\frac{m_{B_d}}{m_{B_s}}
\frac{F^2_{B_d}}{F^2_{B_s}}
\left|\frac{V_{td}}{V_{ts}}\right|^2 r(\mu^+\mu^-),
\end{equation}
where we have introduced the quantities $r(\Delta M)$, $r(\nu\bar\nu)$ 
and $r(\mu^+\mu^-)$ that are all equal unity in models with CMFV. They 
parametrize the deviations from these relations found in several models 
discussed by us below.

Eliminating $|V_{td}/V_{ts}|$ from the three relations above allows 
to obtain three relations between observables that are universal within the
CMFV models. In particular 
from (\ref{dmsdmd}) and (\ref{bmumu}) one finds \cite{Buras:2003td}  
\be\label{R1}
\frac{\mathcal{B}(B_{s}\to\mu^+\mu^-)}{\mathcal{B}(B_{d}\to\mu^+\mu^-)}
=\frac{\hat B_{d}}{\hat B_{s}}
\frac{\tau( B_{s})}{\tau( B_{d})} 
\frac{\Delta M_{s}}{\Delta M_{d}}r, \quad r=\frac{r(\Delta M)}{r(\mu^+\mu^-)}
\ee
that does not 
involve $F_{B_q}$ and consequently contains 
smaller hadronic uncertainties than the formulae considered 
above. It involves
only measurable quantities except for the ratio $\hat B_{s}/\hat B_{d}$
that is now known already from lattice calculations 
with respectable precision \cite{Shigemitsu:2009jy,Laiho:2009eu}
\be\label{BBB}
\frac{\hat B_{s}}{\hat B_{d}}=1.05\pm 0.07, \qquad
\hat B_{d}=1.26\pm0.11, \qquad \hat B_{s}=1.33\pm0.06~.
\ee

Finally one can derive the relations \cite{Buras:2003td} 
\be\label{NonDirect}
 \mathcal{B}(B_q\to\mu^+\mu^-) =  4.36\cdot 10^{-10}\frac{\tau_{B_q}}{\hat B_q}\frac{Y^2(v)}{S(v)}\Delta M_q,
\ee
where $Y(v)$ and $S(v)$ are two master functions of CMFV models \cite{Buras:2003jf} that
are unversal with respect to the flavour 
$(q=d,s,K)$.
The argument $v$ indicates that they depend on specific parameters of a given model. In the SM $v=x_t$. Once $\mathcal{B}(B_q\to\mu^+\mu^-)$ will be measured 
one day, it will be possible to measure the ratio $Y^2/S$ and compare the result
with model predictions of various CMFV models. The important test for CMFV 
will be the same value for $Y^2/S$ obtained from 
$\mathcal{B}(B_s\to\mu^+\mu^-)$ and $\mathcal{B}(B_d\to\mu^+\mu^-)$.
  
The relations (\ref{dmsdmd})-(\ref{R1}), (\ref{NonDirect}) and other 
relations discussed in \cite{Buras:2003jf} can be regarded as {\it 
standard candels of flavour physics}  and the deviations from them may help 
in identifying the correct NP scenario. In particular the parameter $r$ 
in (\ref{R1}) can deviate significantly from unity if
non-MFV sources are present as demonstrated by us
in LHT, RSc and SM4 models.

The relations  in (\ref{NonDirect})
allowed already some time ago to predict $\mathcal{B}(B_{s,d}\to\mu^+\mu^-)$  
in a given CMFV model with substantially smaller hadronic uncertainties 
than found by using directly the formulae 
for the branching ratios in question.
Using the lattice input, in particular (\ref{BBB}), known in 2010, and 
inserting the experimental values of $\Delta M_{d,s}$ allowed to find in the SM \cite{Buras:2010wr}
\be\label{BRSM2}
\mathcal{B}(B_s\to\mu^+\mu^-)_{\rm SM} = (3.2\pm 0.2)\times 10^{-9}, 
\ee
\be\label{BRSM1}
\mathcal{B}(B_d\to\mu^+\mu^-)_{\rm SM} = (1.0\pm 0.1)\times 10^{-10}\,.
\ee

However,  as the uncertainties on the $F_{B_q}$ given in (\ref{lattice1})
have been reduced in 2011 significantly 
it is more appropriate to calculate the branching ratios 
in question directly without using experimental data $\Delta M_{s,d}$.  
The result of this 
calculation is given in  (\ref{LHCb2}) and (\ref{LHCb3}) and
the new values are very close to the 
ones in (\ref{BRSM2}) and (\ref{BRSM1}) obtained two years ago using the 
experimental values of  $\Delta M_{s,d}$. We believe that the results 
given in  (\ref{LHCb2}) and (\ref{LHCb3}),  that have been obtained 
in two ways, are closer to the true values than $3.6\times 10^{-9}$
in the case of $\mathcal{B}(B_s\to\mu^+\mu^-)$  quoted recently by 
some authors.

Let us then confront the CMFV relations listed above 
and predictions with the present data. We observe:

{\bf 1.}
As there are no new CP-violating phases in this framework and 
 formulae in~(\ref{eq:3.43}) apply, CMFV selects solution 1 for $\vub$, i.e. small $\vub$. It should be noted that the small value of 
$S_{\psi\phi}=0.035$ in this framework is fully consistent with 
the LHCb data in (\ref{LHCb1}).

{\bf 2.}
As seen in Table~\ref{tab:SMpred} with small value of $\vub$ the central value of $\varepsilon_K$ in 
the SM is by $16\%$ lower than its experimental value. But in 
CMFV such an enhancement can be naturally obtained simply by increasing 
the value of the one-loop box function $S$. In fact only the increase of $S$ is possible in CMFV \cite{Blanke:2006yh}.

{\bf 3.}
Yet, the enhancement of  $|\varepsilon_K|$ in models with CMFV implies 
automatically in a correlated manner 
enhancements of $\Delta M_d$ and $\Delta M_s$ 
with their ratio unchanged with respect to the SM as 
seen in~(\ref{dmsdmd})\footnote{We assume that CKM parameters $\vub$ 
and $\gamma$ 
have been determined in tree-level decays and as small $\vub$ 
value is also favoured by SM fits, also values of $\vtd$ and 
$\vts$ remain unchanged with respect to the SM.}.
As seen in   Table~\ref{tab:SMpred} the SM values of $\Delta M_{s,d}$ 
are slightly above  the data  and their necessary 
increase required by $|\varepsilon_K|$ worsens the agreement of the 
theoretical values of $\Delta M_{s,d}$ 
 with data significantly even if 
their ratio agrees well with the data.

\begin{figure}[!tb]
 \centering
\includegraphics[width = 0.6\textwidth]{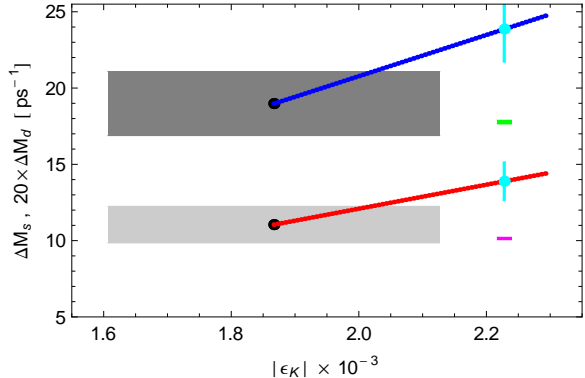}
\caption{$\Delta M_{s}$ (blue) and $20\cdot\Delta M_{d}$ (red) as functions of $|\varepsilon_K|$ in models 
with CMFV for Scenario 1 chosen by these models. The short green and magenta  lines represent the data, while the large black and grey regions the SM 
predictions. More
information can be found in the text.}\label{fig:DeltaMvsepsK}
\end{figure}

We conclude therefore that there is a serious difficulty 
in bringing   $\Delta M_{s,d}$ and $|\varepsilon_K|$
to agree with the data simultaneously within the full class of CMFV models.
In Fig.~\ref{fig:DeltaMvsepsK}  we plot $\Delta M_s$ and $\Delta M_d$ as functions of 
$|\varepsilon_K|$.
In obtaining this plot we have simply varied the 
master one-loop  $\Delta F=2$ function $S$ keeping CKM parameters and other input parameters 
fixed. The value of $S$ at which central experimental value of $|\varepsilon_K|$
is reproduced 
turns out to be $S=2.9$  to        
be compared 
with $S_{\rm SM}=2.31$.
At this value the central values of $\Delta M_{s,d}$ read
\be\label{BESTCMFV}
\Delta M_d=0.69(6)~\text{ps}^{-1},\quad  \Delta M_s=23.9(2.1)~\text{ps}^{-1}~.
\ee
They both differ from experimental values by $3\sigma$.
The error on  $|\varepsilon_K|$ coming dominantly from the 
error of $\vcb$ and the error of the QCD factor $\eta_1$ in the charm contribution 
\cite{Brod:2011ty} is however disturbing. Clearly this plot gives only some 
indication for possible  difficulties of the CMFV and we need 
a significant decrease of theoretical errors in order to see how solid this 
result is.

\begin{figure}[!tb]
 \centering
\includegraphics[width = 0.6\textwidth]{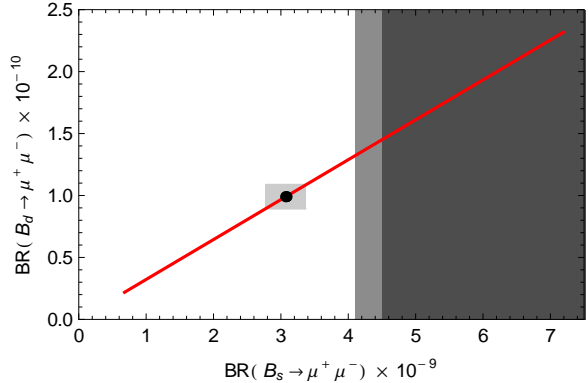}
\caption{ $\mathcal{B}(B_d\to\mu^+\mu^-)$ vs $\mathcal{B}(B_s\to\mu^+\mu^-)$ 
  in models 
with CMFV. SM is represented by the light grey area with black dot. The excluded range by LHCb bound in (\ref{LHCb2}) is in dark and the additional excluded 
grey area corresponds to (\ref{Fleischer}).}\label{fig:BdvsBs}
\end{figure}

{\bf 4.}
Concerning the $B_{s,d}\to\mu^+\mu^-$ decays 
the LHCb upper bound 
on $\mathcal{B}(B_s\to\mu^+\mu^-)$ implies  within CMFV models an upper bound on 
$\mathcal{B}(B_d\to\mu^+\mu^-)$ that is much stronger than the bound in 
(\ref{LHCb3}):
\be\label{boundMFV}
\mathcal{B}(B_{d}\to\mu^+\mu^-) \le 1.3\times 10^{-10}, \quad
({\rm CMFV}),
\ee
where we took into account the correction in~(\ref{Fleischer}).
In Fig.~\ref{fig:BdvsBs} we show $\mathcal{B}(B_d\to\mu^+\mu^-)$ vs $\mathcal{B}(B_s\to\mu^+\mu^-)$ as predicted by CMFV. 
This result and the plot in  Fig.~\ref{fig:DeltaMvsepsK} 
constitute  important tests of 
CMFV. In particular improved data on 
$\mathcal{B}(B_d\to\mu^+\mu^-)$ are very important in this respect. The 
LHCb bound on  $\mathcal{B}(B_d\to\mu^+\mu^-)$ is still outside this plot.

In summary, we find that while in the SM 
$\Delta M_{s,d}$ are consistent with the data and $|\varepsilon_K|$ is 
 visibly below the 
data, a model with CMFV characterized by an enhanced box function 
$S\approx 2.9$, while obtaining the correct value of $|\varepsilon_K|$ 
predicts $\Delta M_{s,d}$ significantly above the data.
As this result depends sensitively on lattice input and the chosen 
$\vub$ and $\gamma$, it will be interesting 
to see whether improved lattice calculations and the tree level 
determination of   $\vub$ and $\gamma$ will confirm our findings 
with higher accuracy. In this context the measurements of all observables 
listed in (\ref{MAINSTARSCP}) and (\ref{MAINSTARSCPC}) will be of crucial 
importance.
Only then we will know
 whether this simplest class of extensions of the SM is ruled out 
and NP contributions 
beyond the CMFV framework are at work. However, if CMFV should 
remain a viable NP scenario  also the experimental branching 
ratio for $\mathcal{B}(B^+\to \tau^+\nu_\tau)$ has to go down by
a factor of 2.

\subsection{Minimal Flavour Violation at Large}

We have already formulated what we mean by CMFV.
 Let us first add here that  the models with CMFV generally contain only
one Higgs doublet and  the top Yukawa coupling dominates.
On the other hand general models with MFV contain more scalar representations, 
in particular two Higgs doublets. Moreover, the operator structure in these 
models can differ from
the SM one. This is the case when  bottom and top Yukawa 
couplings are of comparable size. A well known example is the MSSM with MFV
and large $\tan\beta$. 

In the more general case of MFV the formulation  
with the help of global symmetries present in the limit of vanishing 
Yukawa couplings \cite{Chivukula:1987py,Hall:1990ac} as formulated in 
\cite{D'Ambrosio:2002ex} is elegant and useful.  
Other discussions of various aspects of MFV 
can be found in 
\cite{Colangelo:2008qp,Paradisi:2008qh,Mercolli:2009ns,Feldmann:2009dc,Kagan:2009bn,Paradisi:2009ey,Isidori:2010gz}.

The hypothesis of MFV amounts to assuming that the Yukawas are the only sources of the breakdown of flavour and CP-violation.
The phenomenological implications of the MFV hypothesis formulated in this 
more grander manner than the CMFV formulation given above can be found 
model independently 
by using an effective field theory approach (EFT) \cite{D'Ambrosio:2002ex}. 
In this framework the SM Lagrangian is supplemented by all higher  dimensional
 operators consistent with the MFV hypothesis, built using the Yukawa 
couplings as spurion fields. NP effects in this 
framework
are then parametrized in terms of a few {\it flavour-blind} 
free parameters and SM Yukawa couplings that are solely responsible for 
 flavour violation  and also CP violation if these flavour-blind parameters 
are chosen as {\it real} quantities as done in \cite{D'Ambrosio:2002ex}. 
This approach naturally 
suppresses FCNC processes to the level observed experimentally even in the 
presence of new particles with masses of a few hundreds GeV. It also implies 
specific correlations between various observables, which are not as  stringent 
as in the CMFV but are still very powerful.

Yet, it should be stressed  that the MFV symmetry principle in itself does 
not forbid the presence of
{\it flavour blind} CP violating
sources~\cite{Baek:1998yn,Baek:1999qy,Bartl:2001wc,Paradisi:2009ey,Ellis:2007kb,Colangelo:2008qp,Altmannshofer:2008hc,
Mercolli:2009ns,Feldmann:2009dc,Kagan:2009bn}. Effectively this makes the flavour blind free parameters 
{\it complex} quantities having flavour-blind phases (FBPs). These phases can in 
turn enhance the electric dipole moments {EDMs} of various particles and 
atoms and in the interplay with the CKM matrix can have also profound 
impact on flavour violating observables, in particular the CP-violating ones. 
In the context of the so-called aligned 2HDM model such effects have also been 
emphasized 
in \cite{Pich :2009sp}.

Before turning to a specific model with FBPs let us just mention that in 
this more general framework when FBPs are absent several relations 
of CMFV remain. This is in particular the case of (\ref{bmumu}), where 
$r(\mu^+\mu^-)\approx1$ is found. On the other 
hand (\ref{R1}) can be violated in the presence of new operators that can affect 
$\Delta M_d$ and $\Delta M_s$ differently.

\subsection{${\rm 2HDM_{\overline{MFV}}}$}
\subsubsection{Preliminaries}
We will next discuss a specific 2HDM model, namely  2HDM 
with MFV accompanied by flavour blind CP phases that we will call for 
short ${\rm 2HDM_{\overline{MFV}}}$ \cite{Buras:2010mh} with
 the ``bar'' on MFV indicating the presence of FBPs. 

Let us first list few important points of the ${\rm 2HDM_{\overline{MFV}}}$
 framework.
\begin{itemize}
\item
The presence of FBPs in this MFV framework modifies through their interplay 
with the standard CKM flavour violation the usual characteristic 
relations of  the MFV framework. In particular the mixing induced CP
asymmetries in $B_d^0\to\psi K_S$ and $B_s^0\to\psi\phi$ take the form known 
from non-MFV frameworks like LHT, RSc and 
SM4:
\begin{equation}
S_{\psi K_S} = \sin(2\beta+2\varphi_{B_d})\,, \qquad
S_{\psi\phi} =  \sin(2|\beta_s|-2\varphi_{B_s})\,,
\label{eq:3.43a}
\end{equation}
where $\varphi_{B_q}$ are NP phases in $B^0_q-\bar B^0_q$ mixings.
Thus in the presence of non-vanishing $\varphi_{B_d}$ and $\varphi_{B_s}$, originating here in non-vanishing FBPs, 
these two asymmetries do not measure $\beta$ and $\beta_s$ but $(\beta+\varphi_{B_d})$ and $(|\beta_s|-\varphi_{B_s})$,
respectively.
\item
 The FBPs in the ${\rm 2HDM_{\overline{MFV}}}$ can appear  both 
in Yukawa interactions and  in the Higgs potential. While in  
\cite{Buras:2010mh} only the case of FBPs in Yukawa interactions has been 
considered, in \cite{Buras:2010zm} these considerations have been extended
to include also the FBPs in the Higgs potential.
The two flavour-blind CPV mechanisms can be distinguished
through the correlation between $S_{\psi K_S}$ and $S_{\psi\phi}$ that is
strikingly different if only one of them is relevant. In fact the 
relation between generated new phases are very different in each case:
\be\label{Phase1}
\varphi_{B_d}=\frac{m_d}{m_s}\varphi_{B_s}\quad\quad {\rm and}\quad \varphi_{B_d}=\varphi_{B_s}
\ee
for FBPs in Yukawa couplings and Higgs potential, respectively.
\item 
The heavy Higgs contributions to $\varepsilon_K$ are negligible and 
consequently this model in contrast to CMFV favours the high value 
of $\vub$ for which the SM is consistent with the data on 
$\varepsilon_K$. But this time the presence of the phase $\varphi_{B_d}$ 
allows in principle to remove the $|\varepsilon_K|-S_{\psi K_S}$ anomaly. 
Simultaneously $S_{\psi\phi}$ is enhanced over the SM values with the 
size of enhancement depending on whether FBPs in Yukawas or Higgs potential 
are at work.
\item
The selection of the large value of $\vub$ softens significantly the 
problem with the experimental value of 
$\mathcal{B}(B\to\tau^+\nu_\tau)$ that is by a factor of two 
larger than in the SM.
\item
The branching ratios for $B_{s,d}\to \mu^+\mu^-$ can be sizeably enhanced 
over the SM values but in a correlated manner given by (\ref{bmumu}) with 
$r(\mu^+\mu^-)\approx 1$. 
Moreover, for $S_{\psi\phi}\ge 0.25$ lower bounds on both 
branching ratios are found that are above the SM values and become 
stronger with increasing $S_{\psi\phi}$.
\item
Sizeable FBPs, necessary to explain possible  sizable 
non-standard CPV effects in $B_{s}$ mixing could, in principle, be 
forbidden
by the upper bounds on 
EDMs of the neutron and the atoms. However even for $S_{\psi\phi}=\ord(1)$ 
consistency  with present bounds is obtained 
\cite{Buras:2010zm}.  
\end{itemize}

What is nice about this model is that while having new sources of CP violation 
it has a small number of free parameters
and a  number of definite 
predictions and correlations between various flavour observables 
that provide very important tests of this model. The question then arises 
how this simple model faces most recent LHCb data. Here we only provide 
the first observations. A more detailed study is in progress 
\cite{Buras:2012xxx}.

\begin{figure}[!tb]
 \centering
\includegraphics[width = 0.6\textwidth]{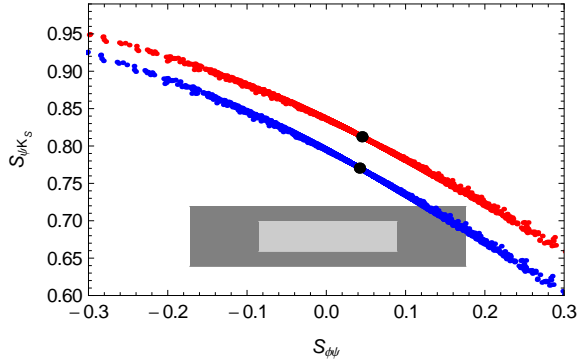}
\caption{ $S_{\psi K_S}$ vs. $S_{\psi \phi}$ in  ${\rm 2HDM_{\overline{MFV}}}$ 
for $\vub=4.0\cdot 10^{-3}$ (blue) and  $\vub=4.3\cdot 10^{-3}$ (red).
  SM is represented by black points while $1\sigma$ ($2\sigma$) experimental 
range by the grey (dark grey) area \cite{Buras:2012xxx}.}\label{fig:SvsS}
\end{figure}

{\bf 1.}
The removal of the $\varepsilon_K-S_{\psi K_S}$ anomaly, which proceeds through 
the  negative phase $\varphi_{B_d}$, is only possible with the help of 
FBPs in the Higgs potential so that optimally 
$\varphi_{B_s}=\varphi_{B_d}$ implying the full dominance of the 
 $Q^{\rm SLL}_{1,2}$ operators as far as CP-violating contributions are 
concerned.

{\bf 2.}
As shown in Fig.~\ref{fig:SvsS} the size of $\varphi_{B_d}$ that is necessary for this removal implies 
in turn $S_{\psi\phi}\ge0.15$ which is 
$2\sigma$ away from 
the LHCb central value in (\ref{LHCb1}). 
Finding in the future that nature chooses a {\it negative} value of 
$S_{\psi\phi}$ and/or small (exclusive) value of $\vub$ would practically 
rule out  ${\rm 2HDM_{\overline{MFV}}}$.

{\bf 3.}
As the CMFV relation (\ref{bmumu}) with $r(\mu^+\mu^-)\approx 1$ 
also applies, 
the 
upper bound on $\mathcal{B}(B_d\to\mu^+\mu^-)$ 
in (\ref{boundMFV})
is  also  
valid in this model.

{\bf 4.}
In the case of the full dominance of NP effects from the Higgs potential, 
represented by the operators  $Q^{\rm SLL}_{1,2}$,
 also 
in the case of $\Delta M_{s,d}$
the CMFV relation (\ref{dmsdmd}) with $r(\Delta M)\approx 1$ applies 
and consequently 
(\ref{R1})  with $r=1$ is valid. 
 Yet, the CMFV correlation between 
$\varepsilon_K$ and $\Delta M_{s,d}$ is absent and $\Delta M_{s,d}$ can 
be both suppressed and enhanced if necessary. 
The predicted value of $\Delta M_s/\Delta M_d$ is close to the one found 
in the data unless large contributions of operators  $Q^{\rm LR}_{1,2}$ 
are present. They suppress $\Delta M_s$ with basically no effect on 
$\Delta M_d$. Then $r(\Delta M)\ge 1$ and consequently for this 
range of parameters $r\ge 1$. Thus at first sight this model provides 
a better description of $\Delta F=2$ data than the SM and models with 
CMFV. A more definite statement will be provided soon \cite{Buras:2012xxx}.

{\bf 5.}
As already seen in the plots in \cite{Buras:2010zm} 
in the range of  the presently allowed values for $S_{\psi\phi}$ there is basically no 
correlation between this CP-asymmetry and $\mathcal{B}(B_{q}\to\mu^+\mu^-)$. 
Consequently the latter, even if already rather constrained by the data 
can be still smaller or larger than the SM values. More work is needed 
to provide more definite statements \cite{Buras:2012xxx}.

We are looking forward to improved experimental data and improved lattice 
calculations to find out whether this simple model can satisfactorily describe 
the observables considered by us.

\subsection{Littlest Higgs Model with T-parity}
We will next discuss two models having the operator structure of the 
SM but containing new sources of flavour and CP-violation. This is the Littlest 
Higgs Model with T-parity (LHT) and the SM4, the SM extended by a fourth sequential 
generation of quarks and leptons.

The Littlest Higgs model
without \cite{ArkaniHamed:2002qy} T-parity 
has been invented to solve the problem of the quadratic 
divergences in the Higgs mass without using supersymmetry.
In this approach the cancellation of divergences in $ m_H$ is achieved with 
the help of new particles of the same spin-statistics. 
Basically the SM Higgs is kept light 
because it is a pseudo-Goldstone boson of a spontaneously broken global 
symmetry:
\be 
SU(5)\to SO(5).
\ee
Thus the Higgs is protected from acquiring a large mass by a global symmetry, 
although in order to achieve this the gauge group has to be extended
to
\be
 G_{\rm LHT}=SU(3)_c\times [SU(2)\times U(1)]_1\times[SU(2)\times U(1)]_2
\ee
 and 
the Higgs mass generation properly arranged 
({\it collective symmetry breaking}). 

In order to make this model 
consistent with electroweak precision tests and simultaneously having
the new particles of this model in the reach of the LHC, a discrete symmetry,
T-parity, has been introduced \cite{Cheng:2003ju,Cheng:2004yc}. 
Under T-parity all SM particles are {\it even}.
Among the new particles only a heavy $+2/3$ charged $T$ quark belongs to the
even sector. Its role is to cancel the quadratic divergence in the Higgs
mass generated by the ordinary top quark. The even sector and also the 
model without T-parity (LH model) belong to the CMFV class if only flavour violation 
in the down-quark sector is considered \cite{Buras:2004kq,Buras:2006wk}. 
But it should be stressed that the T-even sector of the LHT model differs 
from 
the LH model considered in the latter papers where also loop diagrams with 
heavy gauge bosons contribute.
The high masses of new particles implied by electroweak precision tests in 
the LH model allow only for small deviations from the SM. In the LHT model
all NP effects in the T-even sector come from the T-quark interactions 
with standard quarks mediated by the SM gauge bosons and the effects can be in principle larger.

Yet,  from the point of view of FCNC processes more interesting is the T-odd
sector. Because of T-parity it contains first of all 
three doublets of heavy mirror quarks and three doublets
of mirror leptons that correspond to the SM fermions and communicate with the 
latter by means of heavy 
$W^\pm_H$, $Z_H^0$ and $A^0_H$ gauge bosons that are also odd under 
T-parity. These interactions are 
governed by new mixing matrices that bring in new flavour parameters, in 
particular new CP phases \cite{Hubisz:2005bd,Blanke:2006xr}. The T-parity 
partner of the $T$ quark does not play any role in FCNC processes. It should 
be observed that in the limit of degenerate mirror quark masses T-odd sector 
does not contribute to FCNC processes and the LHT model represented then in 
these processes only by T-even sector belongs to the class of CMFV 
models.  The problems of CMFV models identified before indicate that the
T-odd sector is crucial for this model to achieve the agreement with data.

Let us summarize the main implications of this model for flavour 
phenomenology which are based on \cite{Blanke:2006sb,Blanke:2006eb,Blanke:2007db,Blanke:2009pq,Blanke:2009am}. The plots given in these papers allow 
to answer at least qualitatively 
how this model faces the recent LHCb data. Fortunately our new analysis 
triggered by new LHCb data allows to see  the present status of the 
flavour physics in the LHT model even at a quantitative level 
\cite{Blanke:2012yyy}:

{\bf 1.}
The difference between the CMFV models and the LHT model, originating in 
the presence of mirror quarks and new mixing matrices, is the violation 
of the usual CMFV relations between $K$, $B_d$ and $B_s$ 
systems of which we have shown some above. This allows to remove 
the $\varepsilon_K-S_{\psi K_S}$ anomaly for both scenarios of $\vub$ and 
also improve agreement with $\Delta M_s$ and $\Delta M_d$. As this can 
be done simultaneously, the LHT model provides a better description 
of the data than the CMFV models.

{\bf 2.}
 Interestingly, in this model it was not possible to obtain 
$S_{\psi\phi}$ of $\ord(1)$ and  
 values above 0.3 were rather unlikely. The LHCb result in (\ref{LHCb1}) 
can therefore be considered as a relief for this model 
in which also negative values for
$S_{\psi\phi}$ as opposed to  ${\rm 2HDM_{\overline{MFV}}}$ are possible.

{\bf 3.}
$\mathcal{B}(\klpn)$ and $\mathcal{B}(\kpn)$ can be enhanced up to factors of
 3 and 2.5, respectively but not simultaneously with $S_{\psi\phi}$. 
Therefore the small values for $S_{\psi\phi}$ found by LHCb are good news 
for $\klpn$ and $\kpn$ in the LHT model, although large enhancements of 
their branching ratios although possible are not guaranteed.

{\bf 4.}
 Rare $B$-decays turn out to be SM-like but still some enhancements 
are possible. 
In particular $\mathcal{B}(B_{s}\to\mu^+\mu^-)$ is predicted to be 
larger than its SM value but it can only be enhanced by
 $30\%$, where a
significant part of this enhancement comes from the T-even sector.
The effects in  $\mathcal{B}(B_{d}\to\mu^+\mu^-)$ can be larger and 
also suppression is possible.

The plots presented in \cite{Blanke:2012yyy} should facilitate monitoring 
the future confrontations of the LHT model with the data 
and to find out whether this simple model can satisfactorily describe 
the observables considered by us.
Further phenomenological discussions of LHT model can be found in 
original papers quoted above and in \cite{Goto:2008fj,delAguila:2008zu}.

\subsection{The SM with Sequential Fourth Generation}
One of the simplest extensions of the SM3
is the addition of a sequential fourth generation (4G)
of quarks and leptons \cite{Frampton:1999xi}
{(hereafter referred to as SM4)}. Therefore it is of interest to study 
its phenomenological implications. 
Beyond flavour physics possibly the 
most interesting implications of the presence of 4G would be 
the viability of electroweak baryogenesis \cite{Hou:2008xd,Kikukawa:2009mu,Fok:2008yg} and 
dynamical breakdown of electroweak symmetry triggered by the presence of 
4G quarks
 \cite{Holdom:1986rn,Hill:1990ge,King:1990he,Burdman:2007sx,Hung:2009hy,Hung:2009ia,Holdom:2010za,Hashimoto:2010fp}.

Yet, the LHC data indicate that our nature seems to have only three sequential 
generations of quarks and leptons, although the story of SM4 is not over yet.
Selected recent papers demonstrating that SM4 is in trouble 
outside the flavour physics can be found in \cite{Djouadi:2012ae,Kuflik:2012ai}.

During the last ten years a number of flavour analyses of SM4 have been 
performed
\cite{Arhrib:2002md,Hou:2005yb,Hou:2006mx,Soni:2008bc,Soni:2010xh,Herrera:2008yf,Bobrowski:2009ng,Eberhardt:2010bm,Eilam:2009hz,Buras:2010pi,Buras:2010nd,Buras:2010cp,Hou:2008yb,Das:2010fh}.
The SM4
introduces three
new mixing angles $s_{14}$, $s_{24}$, $s_{34}$ and two new phases in the 
quark sector and can still have a significant impact on flavour 
phenomenology. Similarly to the LHT model it does not introduce any 
new operators but brings in new sources of flavour and CP violation 
that originate in the interactions of the four generation fermions 
with the ordinary quarks and leptons that are mediated by the SM 
electroweak gauge bosons. Thus in this model, as opposed to the LHT 
model, the gauge group is the SM one. This implies smaller number of 
 free parameters.

An interesting virtue of the SM4 model is the non-decoupling of new 
particles. Therefore, unless the model has non-perturbative Yukawa 
interactions, the 4G fermions are bound to be observed at the LHC with 
masses below $600\gev$. This did not happen yet. In spite of this 
it is of interest to see what is the impact of the recent LHCb data on the 
correlations found in our 
analyses of $K$ and $B$ flavour 
physics \cite{Buras:2010pi,Heidsieck:2010ue}.

In what follows we list
the most interesting patterns of quark flavour 
violation in the SM4 we have found in these papers and indicate how they are 
modified through LHCb at a qualitative level.

{\bf 1.}
As before the presence of new sources of flavour violations allows to 
solve all existing tensions related to $\Delta F=2$ observables.

{\bf 2.}
We have also found that the desire to explain large values of $S_{\psi\phi}$ 
signaled in 2010
implies
 uniquely the suppressions of 
 the CP asymmetries $S_{\phi K_S}$ and $S_{\eta^\prime K_S}$ below their 
SM values that are equal
to $S_{\psi K_S}$. Such suppressions were still visible in the
  2010 data.
 This correlation has been pointed out in \cite{Hou:2005yb,Soni:2008bc},  
 however we observed that for $S_{\psi\phi}$ significantly larger
 than 0.6 the values of $S_{\phi K_s}$ and $S_{\eta' K_s}$ are below their central values indicated by the data, although some non-perturbative uncertainties
 were involved here. With the new value for  $S_{\psi\phi}$ in (\ref{LHCb1}) 
 this suppression is absent and the values of  $S_{\phi K_S}$ and $S_{\eta' K_S}$ 
 in SM4 are compatible with the value of $S_{\psi K_S}$ as meanwhile also
 seen in the data.

{\bf 3.}
The enhanced value of $S_{\psi\phi}$ would imply a sizable enhancement of $\mathcal{B}(B_s\to \mu^+\mu^-)$
over the SM3 prediction although this effect is much more modest than 
in SUSY models where the Higgs penguin  with large $\tan\beta$ is at work. Yet,
values as high as $8\cdot 10^{-9}$ were certainly possible in the SM4 in 2010. 
On the other hand large values of $S_{\psi\phi}$ would preclude 
non-SM values of $\mathcal{B}(B_d\to\mu^+\mu^-)$. Consequently the CMFV relations 
in (\ref{bmumu}) and (\ref{R1}) can be strongly violated in this model.
The small value of $S_{\psi\phi}$ and the stringent upper bound on 
$\mathcal{B}(B_s\to\mu^+\mu^-)$ from LHCb implies now that  $\mathcal{B}(B_d\to\mu^+\mu^-)$ can significantly depart from the SM
value. On the other hand $\mathcal{B}(B_s\to\mu^+\mu^-)$ is 
SM-like with values {\it below} SM prediction being more likely than above it.
In any case the deviations from  CMFV relations 
in (\ref{bmumu}) and (\ref{R1}) could still be sizable. 
All these features are clearly seen in the plots of our 2010 paper 
\cite{Buras:2010pi} and can be considered as predictions of SM4 found 
prior to the LHCb results. See also Fig.~\ref{fig:BsmumubsBdmumu}.

{\bf 4.}
Possible enhancements of $\mathcal{B}(\kpn)$ and $\mathcal{B}(\klpn)$ over the SM3 values are still possible as these branching
ratios were not strongly 
correlated with $S_{\psi\phi}$ and $B_{s,d}\to\mu^+\mu^-$.
 Even in the presence of SM-like values for $S_{\psi\phi}$ and $\mathcal{B}(B_s\to \mu^+\mu^-)$, significant effects in the other
decays like 
$K_L\to\pi^0\ell^+\ell^-$ and $K_L\to\mu^+\mu^-$ are possible.

{\bf 5.}
We have found that for large positive values of $S_{\psi\phi}$ the predicted value of $\epe$ is significantly below the data, unless the
 hadronic matrix elements of the electroweak penguins are sufficiently suppressed with respect to the large $N$ result and the
ones of QCD penguins enhanced.
With the small value of  $S_{\psi\phi}$ from LHCb this is not an issue anymore.

In summary, provided the four generation quarks will still be found, our 
qualitative discussion shows that if present, the new quarks can still 
have a potential impact on quark flavour physics. The same comment applies 
to new heavy leptons \cite{Buras:2010cp}.

\subsection{Supersymmetric Flavour Models (SF)}
None of the 
supersymmetric particles 
has been seen so far. However,
one of the important predictions of 
the 
simplest realization of this scenario, the MSSM with $R$-parity, 
is a light Higgs with $m_H\le 130\gev$. The events at the LHC around 
$125\gev$ could indeed be the first hints for a Higgs of the MSSM but 
it will take some time to verify it. In any case MSSM remains still a viable 
NP scenario at scales $\ord(1\tev)$.

Concerning the FCNC processes  
squarks, sleptons, gluinos, charginos,
 neutralinos, charged Higgs particles $H^{\pm}$ and additional heavy
 neutral scalars  
can contribute to FCNC transitions through box and 
penguin diagrams. 
New sources of flavour
and CP violation come from the misalignment of quark and squark mass 
matrices
 and similar new flavour and CP-violating effects are present in the lepton
sector. Some of these effects can be strongly enhanced at large $\tan\beta$
and the corresponding observables provide stringent constraints on
the parameters of the MSSM.
In particular $B_{s,d}\to \mu^+\mu^-$ can be strongly enhanced
and the CP asymmetry $S_{\psi\phi}$ can be $\ord(1)$.

The SUSY dreams of large $\mathcal{B}(B_s\to\mu^+\mu^-)$ and $S_{\psi\phi}$ 
have not been realized at the LHCb and the data from LHCb listed in 
(\ref{LHCb1}), (\ref{LHCb2}) and (\ref{LHCb3}) have certainly an impact on
SUSY predictions. 

In what follows we will only make a first look at the impact of the
LHCb data on the supersymmetric 
flavour (SF) models having 
flavour symmetries that
allow for  some understanding of 
the flavour structures in the Yukawa couplings and in SUSY soft-breaking 
terms, adequately suppressing FCNC and CP violating phenomena and solving
SUSY flavour and CP problems.

The SF models can be divided into two broad
classes depending on whether they are based on abelian or non-abelian 
flavour symmetries. Moreover, their phenomenological output crucially
depends on whether the flavour and CP violations are governed by 
left-handed (LH)
currents or if there is an important new right-handed (RH) current component
 \cite{Altmannshofer:2009ne}.

In \cite{Altmannshofer:2009ne} we have
performed an extensive study of processes governed by $b\to s$ transitions 
in the SF models and of their correlations with processes governed by 
$b\to d$ transitions, 
$s\to d$ transitions, $D^0-\bar D^0$ mixing, LFV 
decays, electric dipole moments and $(g-2)_{\mu}$. 
Both abelian and non-abelian flavour models have been considered as well as the
flavour blind MSSM (FBMSSM) and the MSSM with MFV. It has been shown how
 the characteristic patterns of correlations among the considered flavour 
observables allow to distinguish between these different SUSY scenarios and 
also to distinguish them from RSc and LHT scenarios of NP.

Of particular importance in our study were the correlations between 
the CP asymmetry $S_{\psi\phi}$ and
$B_s\rightarrow\mu^+\mu^-$, between the observed anomalies in 
$S_{\phi K_s}$ and $S_{\psi\phi}$, between 
$S_{\phi K_s}$ and $d_e$, between $S_{\psi\phi}$ and $(g-2)_{\mu}$ and 
also those involving LFV decays.

In the context of our study of the SF models we have analysed the 
following representative scenarios:
\begin{itemize}
\item [1.] Dominance of RH currents 
(abelian model by Agashe and Carone (AC)\cite{Agashe:2003rj}),
\item [2.] Comparable LH and RH currents with CKM-like mixing
  angles represented by the special version (RVV2) 
of the non abelian $SU(3)$ 
model by
Ross, Velasco and Vives \cite{Ross:2004qn} as discussed in 
\cite{Calibbi:2009ja}. 
\item [3.] In the second non-abelian $SU(3)$ 
 model by Antusch, King and Malinsky (AKM) \cite{Antusch:2007re} analyzed 
by us the RH contributions are CKM-like but 
 new LH contributions in contrast to the RVV2 model can be 
suppressed arbitrarily at the high scale. Still 
they can be generated by RG effects at lower scales. To first approximation 
the version of this model considered by us can be characterized by NP being
dominated by CKM-like RH currents.
\item [4.] Dominance of CKM-like LH currents in non-abelian 
models~\cite{Hall:1995es}.
\end{itemize}

The distinct patterns of flavour violation found in each scenario have
been illustrated with several  plots that can  be found
in figures 11-14 of   \cite{Altmannshofer:2009ne}. The power of these 
plots lies in the fact that even without a detailed numerical 
analysis one can on a qualitative level state what is the impact 
of the LHCb data on these results. A more quantitative analysis 
would require full numerical analysis taking also collider constraints 
on the masses of new particles involved. Before the situation with SUSY 
searches at the LHC is settled down such an analysis appears premature 
to us. Keeping this in mind we 
will just recall some of 
the main messages from \cite{Altmannshofer:2009ne}, stating how at first 
sight they are modified by the new data.

{\bf 1.}
Supersymmetric models with RH currents (AC, RVV2, AKM) and those with 
exclusively LH currents
can be in principle globally distinguished by the values of the CP-asymmetries 
$S_{\psi\phi}$ and $S_{\phi K_S}$ with the following important result: 
none of
the models considered in \cite{Altmannshofer:2009ne} could simultaneously explain the $S_{\psi\phi}$ and
$S_{\phi K_S}$ anomalies observed in the data in 2009. 
In the models with RH currents,
$S_{\psi\phi}$ can naturally be much larger than its SM value, while 
$S_{\phi K_S}$ remains either SM-like or its correlation with $S_{\psi\phi}$ 
is inconsistent with the data. 
On the contrary, in the models with LH currents only,
$S_{\psi\phi}$ remains SM-like, while the  $S_{\phi K_S}$  anomaly could  
easily be explained in 2009.
The data on
$S_{\psi\phi}$ and $S_{\phi K_S}$ in 2012 do not indicate any potential 
anomalies in both observables and the distinction between these 
two classes of models on the basis of these two asymmetries will be difficult 
unless $S_{\psi\phi}$ is finally found visibly different from the SM value. This could 
only be explained in models with RH currents.

{\bf 2.}
The desire to explain large values of $S_{\psi\phi}$ in 2009 within the models with
RH currents unambiguously implied, in the case of the AC and the AKM models,
values of
$\mathcal{B}(B_s\to\mu^+\mu^-)$ as high as several $10^{-8}$. In the 
RVV2 model such values were also possible but not necessarily implied
by the large value of $S_{\psi\phi}$.  With the new range for  $S_{\psi\phi}$ a lower 
bound on $\mathcal{B}(B_s\to\mu^+\mu^-)$ slightly above the SM value in  AC and the AKM models is still present for
$|S_{\psi\phi}|\approx 0.2$.
It should also be emphasized that all these models can provide negative 
values of $S_{\psi\phi}$ which is not possible in ${\rm 2HDM_{\overline{MFV}}}$.

\begin{figure}[!tb]
 \centering
\includegraphics[width = 0.8 \textwidth]{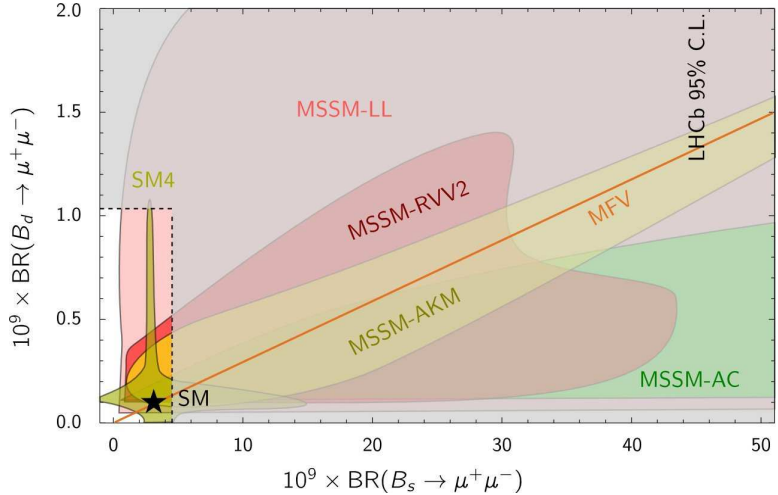}
\caption{Results in different SF models  \cite{Altmannshofer:2009ne} as collected in \cite{Straub:2010ih}. The impact of the new
LHCb bounds in 
(\ref{LHCb2}) and (\ref{LHCb3}) is shown.}\label{fig:BsmumubsBdmumu}
\end{figure}

{\bf 3.}  As seen in Fig.~\ref{fig:BsmumubsBdmumu} a better distinction 
between different SF models is offered by 
the ratio 
$\mathcal{B}(B_d\to\mu^+\mu^-)/\mathcal{B}(B_s\to\mu^+\mu^-)$ that in the AC and RVV2  models turns out to be
dominantly below its MFV prediction (straight line) and could be much smaller than the latter.
In the AKM model this ratio stays much closer to the MFV value of roughly 
$1/32$  \cite{Buras:2003td,Hurth:2008jc} and can  be smaller or larger than 
this value with equal probability.
Interestingly,  in the LH-current-models, the ratio 
$\mathcal{B}(B_d\to\mu^+\mu^-)/\mathcal{B}(B_s\to\mu^+\mu^-)$ can not only deviate
significantly from its MFV value of approximately $1/32$, 
but in contrast to the models with 
RH currents considered by us can also be much larger than the latter value. 
Consequently,
$\mathcal{B}(B_d\to\mu^+\mu^-)$ as high as $1\times 10^{-9}$ is still 
possible, saturating the present upper bound in (\ref{LHCb3}) while being consistent with the bounds on
$\mathcal{B}(B_s\to\mu^+\mu^-)$ in (\ref{LHCb2}). Evidently the new 
LHCb data had a significant impact on the ratio in question and 
from the present 
perspective there is more room for NP contributions dominated by LH currents.

{\bf 4.}
Next, while the abelian AC model resolves the present UT tensions 
\cite{Lunghi:2008aa,Buras:2008nn,Lunghi:2009sm,Buras:2009pj,Lunghi:2009ke,Lunghi:2010gv,UTfit-web,Lenz:2010gu} 
through the modification of the ratio $\Delta M_d/\Delta M_s$, the 
non-abelian flavour models RVV2 and AKM provide the solution through
NP contributions to $\varepsilon_K$. As the ratio $\Delta M_d/\Delta M_s$ 
within the SM is roughly correct and cannot be changed by much, it appears 
at first sight that the AC model cannot remove the $|\varepsilon_K|-S_{\psi K_S}$  anomaly. However, in order to be sure a new analysis of this model has 
to be performed.

{\bf 5.}
The branching ratios for 
$K\to\pi \nu\bar\nu$ decays in the supersymmetric models considered by us
remain SM-like and can be distinguished from RSc and LHT models where  
they can still be significantly enhanced.

 In summary although the large range of departures from SM expectations 
 found in \cite{Altmannshofer:2009ne} has been significantly narrowed, 
 still significant room for novel SUSY effects is present in quark 
 flavour data. Assuming that SUSY particles will be found, the future 
improved data for $B_{s,d}\to\mu^+\mu^-$ and $S_{\psi\phi}$ as well as
$\gamma$ combined with $\vub$ should help in distinguishing between various 
supersymmetric flavour models.

\subsection{Supersymmetric SO(10) GUT model}

GUTs open the possibility to transfer the neutrino mixing matrix $U_\text{PMNS}$ to the quark sector. This is accomplished in a
controlled way in a SUSY GUT model proposed by Chang, Masiero and Murayama (CMM model) where the atmospheric neutrino mixing
angle induces new $b\to s$ and $\tau\to \mu$ transitions \cite{Moroi:2000tk,Chang:2002mq}. We have performed a global analysis in
the CMM model including an extensive renormalisation group (RG) analysis to connect Planck-scale and low-energy parameters 
\cite{Girrbach:2011an,Girrbach:2011wt}. Since this work has been done before the new LHCb data on $S_{\psi\phi}$ and
$\mathcal{B}(B_s\to \mu^+\mu^-)$ we want to comment on the implications of these data for this model.

The basic properties of this model can be summarized as follows:  The flavour symmetry which is exact at the Planck scale
is broken at the SO(10) scale which manifests itself in the appearance of a non-renormalizable operator in the SO(10) superpotential. As
a consequence we get a natural hierarchy between top and bottom Yukawa couplings which finally leads to small $\tan\beta$ between
3 and 10 (as opposed to other SO(10) GUT models with $y_t \approx y_b$ and $\tan\beta \approx 50$). While at $M_\text{Pl}$ the
soft masses are universal, we get a large splitting between the masses  of the 1st/2nd  and 3rd down-squark and
charged-slepton generation  at the electroweak scale due to RG effects of $y_t$. The SO(10) symmetry is broken down to the SM gauge
group via SU(5) where the
right-handed down quarks and the lepton doublet are unified in the $\mathbf{\bar 5}$. 
Then not only the neutrinos are rotated with
the PMNS matrix, but the whole $\mathbf{\bar{5}}$-plet and the corresponding supersymmetric partners. In addition, a  model
parameter~-- a CP violating phase $\xi$~-- enters the rotation of right-handed down (s)quarks.
 Consequently the neutrino
mixing angles are transferred to the right-handed down-squark/charged-slepton sector which then induces $b\to s$ and
$\tau\to\mu$ transitions and CP violation in \bbms{} via SUSY loops. The flavour effects in the
CMM model are mainly determined by the generated mass splitting and the structure of the PMNS matrix.

Whereas effects in \kkm, \bbmd{} and $\mu\to e\gamma$ are very small in the original version of the model, large contributions are
predicted in
observables connecting the 2nd and 3rd generation. However, since the CMM model at low energies appears as a special version of
the
MSSM with small $\tan\beta$, effects in $B_s\to\mu^+\mu^-$ are negligible such that this branching ratio stays SM-like consistent
with the recent upper bound from LHCb. In \cite{Girrbach:2011an} we therefore focused on $b\to s\gamma$, $\tau\to\mu\gamma$,
$\Delta M_s$ and $S_{\psi\phi}$, where e.g. $\mathcal{B}(\tau\to\mu \gamma)$ alone puts a lower bound on $M_{\tilde q}$ (mass of the 1st/2nd
squark generation). Here we will
concentrate on $\Delta M_s$ and $S_{\psi\phi}$:

\begin{figure}[!tb]
 \centering
\includegraphics[width = 0.7\textwidth]{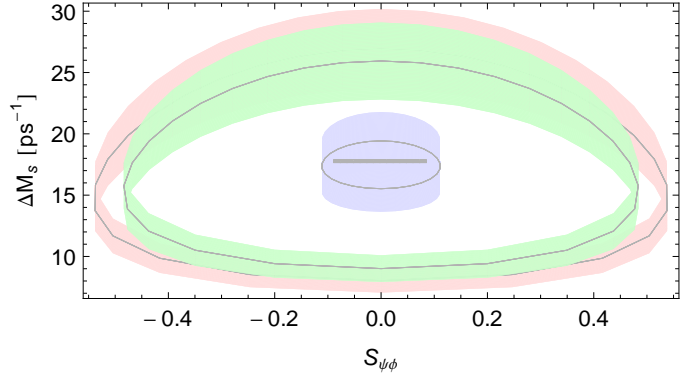}
\caption{Correlation between $\Delta M_s$ and $S_{\psi\phi}$ for scenario~1 (blue), 2 (green), 3 (red) (from inside to outside).
The width of the coloured bands comes from the error of $F_{B_s}^2 \hat{B}_{B_s}$ (the solid grey line corresponds to the central
value).
}\label{fig:CMMDMsvsSinphis}
\end{figure}

In the CMM model two operators contribute to \bbms: $Q_1^\text{VLL}$ and $Q_1^\text{VRR}$. The corresponding Wilson coefficient
can then be written as $C = C_L + e^{-2i\xi}|C_R^\text{CMM}|$ which makes clear that there can be new CP violating effects in
$S_{\psi\phi}$. In view of the data from CDF and D0 on $S_{\psi\phi}$ this property was very welcomed in 2010.  As an example, we
show 
in Fig.~\ref{fig:CMMDMsvsSinphis}  the correlation between  $\Delta M_s$ and 
$S_{\psi\phi}$ for three selected points  in the
GUT parameter space that are consistent
with $b\to s\gamma$ and $\tau\to \mu\gamma$, namely for  $\tan\beta = 7$, $\text{arg}(\mu) = 0$ and
\begin{enumerate}
  \item  $M_{\tilde q}= 1500~$GeV, $m_{\tilde g} = 900~$GeV, $a_1^d/M_{\tilde q} = 1.5$, 
(blue)
 \item  $M_{\tilde q}= 1500~$GeV, $m_{\tilde g} = 600~$GeV, $a_1^d/M_{\tilde q} = 1.5$,
(green)
 \item  $M_{\tilde q}= 2000~$GeV, $m_{\tilde g} = 700~$GeV, $a_1^d/M_{\tilde q} = 1.8$,
(green)
\end{enumerate}
 while simultaneously scanning over $\xi\in[0,\pi]$. As one can see it is easily possible to get large CP violation in \bbms,
such
  that scenario~2 and~3 were consistent with the experimental values of $\Delta M_s$ and the large $S_{\psi\phi}$ found in CDF
  and D0. However, now only scenario 1 is consistent with the new LHCb measurement of $S_{\psi\phi}$ and  this implies new constraints
  on the model parameters, e.g. on the ratio $m_{\tilde g}/M_{\tilde q}$. Consequently one previous advantage of the CMM model
  over mSUGRA/CMSSM scenarios~-- the ability to generate a large $S_{\psi\phi}$~-- is now gone. 

The authors of \cite{Trine:2009ns} studied the implications of corrections to the unification of down-quark and charged-lepton
Yukawa couplings $\mathsf{Y}_d = \mathsf{Y}_\ell$ in the CMM model. This relation works remarkably well for the third generation
but not for the two lighter ones. Therefore one has to include corrections that are  generated by higher-dimensional
Yukawa operators suppressed by powers of $M_\text{Pl}$ which do not spoil the successful bottom-tau unification but can a priori
have arbitrary flavour structure. Consequently these operators have implications for \kk and \bbmd{} but do not change the CMM
predictions for $b\to s$ and $\tau\to\mu$ transitions.  One main result of \cite{Trine:2009ns}  is that the flavour structure of
the higher-dimensional Yukawa operators is very much  constrained due to $|\varepsilon_K|$ which means that the dimension-5-Yukawa
couplings and
tree level Yukawa couplings must be nearly aligned. A similar result using $\mu\to e\gamma$ was found in \cite{Girrbach:2009uy}.
 In \cite{Trine:2009ns} it was also shown that the tension in the SM between $\sin2\beta$ predicted from $|\varepsilon_K|$ and
$\Delta M_s/\Delta M_d$, and its direct measurement from $S_{\psi K_S}$ can be removed with the help of higher-dimensional Yukawa
couplings. CMM effects can then appear either in the UT side $R_t$ through contributions from $\Delta M_s$ or both in $R_t$
and $S_{\psi K_S}$ through an additional phase.

The CMM model can still serve as an alternative benchmark scenario to the popular constraint MSSM. It has only seven input
parameters, is
universal at $M_\text{Pl}$ and not at $M_\text{GUT}$ as in the CMSSM, it has a very clear flavour structure and in contrast to
the CMSSM  hadronic and leptonic observables are related.

\subsection{ Supersymmetric $SU(5)$ GUT with RH  neutrinos (RN): 
$SSU(5)_{\rm RN}$}

We will next consider a supersymmetric $SU(5)$ GUT enriched by 
 right-handed neutrinos ($SSU(5)_{\rm RN}$) accounting for the neutrino
 masses and mixing angles by means of a type-I see-saw mechanism.
Since SUSY-GUTS generally predict FCNC and CP violating
processes to occur both in the leptonic and hadronic sectors,
 we have performed in \cite{Buras:2010pm} an
extensive study of FCNC and CP violation in both sectors, analyzing possible
hadron/lepton correlations among observables. In particular, we have monitored
how in this framework the tensions observed in the UT analysis can be 
resolved. 
Here
the correlations between leptonic and hadronic processes taking place 
between the same
generations like $\mu\to e\,\gamma$ and $s\to d$ or  $\tau\to\mu\,\gamma$ and $b \to s$ transitions exist.

The main results of our study of the  $s\to d$ transitions and of 
their correlations with the $\mu\to e$ transitions remains basically unchanged 
except that the effects are likely to be smaller as the lower bounds on 
supersymmetric particles increased. We refer to  \cite{Buras:2010pm} 
for details. Here we concentrate on the impact of LHCb data on selected 
results of
our study of the $b\to s$ transitions and of
their correlations with the  $\tau\to\mu$ transitions. They are:

{\bf 1.} Non-standard values for $S_{\psi\phi}$ implied in 2010 
a lower bound for 
$\mathcal{B}(\tau\to\mu\gamma)$ within the reach of SuperKEKB and SuperB.
 However, we also found that the $(g-2)_\mu$ 
anomaly can be solved only for large $\tan\beta$
values where we found $|S_{\psi\phi}|\leq 0.2$ for $\Delta a^{\rm SUSY}_{\mu}\geq 1\times 10^{-9}$
while being still compatible with the constraints from 
$\mathcal{B}(\tau\to\mu\gamma)$. Now even if our analysis implied already at that 
time only moderate values of   $|S_{\psi\phi}|\leq 0.2$, the fact that 
large values of  $\tan\beta$ are not welcome anymore in view of the LHCb
upper bound on $\mathcal{B}(B_s\to\mu^+\mu^-)$ changes the analysis 
of  $(g-2)_\mu$. It appears that  SUSY models do not provide the explanation 
for this anomaly anymore \cite{Jegerlehner:2012ju}.

{\bf 2.}
We recall that in this model
$S_{\psi K_S}$ remains SM-like to a very good extent and consequently the 
solution of the UT anomalies by means of CPV effects in $b\to d$ mixing 
is not possible. 
However,
the UT anomalies can be solved by means of a negative NP contribution 
to  $\Delta M_d/\Delta M_s$,
implying a lower bound for $\mathcal{B}(\tau\to\mu\gamma)$ within the 
reach of  SuperKEKB and SuperB  and large
values for the angle $\gamma$. This scenario will be probed or falsified 
in due time at the LHCb through a precise tree level measurement of the 
latter UT 
angle. However already now it appears to us that the modification of 
 $\Delta M_d/\Delta M_s$, through the increase of $\gamma$ is not favoured 
by the present data and the model may have a problem similarly to 
the AC supersymmetric flavour model  in removing the anomalies. 

{\bf 3.} Both $\mathcal{B}(B_s\to\mu^+\mu^-)$ and $\mathcal{B}(B_d\to\mu^+\mu^-)$ can reach
large non-standard values. While for  $\mathcal{B}(B_s\to\mu^+\mu^-)$ this is no 
longer possible, this could still be the case for  $\mathcal{B}(B_d\to\mu^+\mu^-)$.
In such a case sizable departures from the MFV prediction
$\mathcal{B}(B_s\to\mu^+\mu^-)/\mathcal{B}(B_d\to\mu^+\mu^-)\approx |V_{ts}/V_{td}|^2$ will 
be present and as our analysis shows this could allow 
enhanced values of $\mathcal{B}(\tau\to\mu\gamma)$, possibly in the 
reach of  SuperKEKB and SuperB.

\subsection{The flavour blind MSSM (FBMSSM)}
The flavour blind MSSM (FBMSSM) scenario \cite{Baek:1998yn,Baek:1999qy,Bartl:2001wc,Ellis:2007kb,Altmannshofer:2008hc} having new FBPs in the soft sector
belongs actually to the class of MFV models or even better is
 of $\overline{\rm MFV}$ type.
However it is a supersymmetric framework and
 we mention this model here.

The FBMSSM has fewer parameters than the general MSSM
 implying striking correlations
between various observables analyzed in  \cite{Altmannshofer:2008hc}. 
Here we only make a few remarks on the $\Delta F=2$ observables as 
the recent LHCb results are actually good news  for this model,  even 
if the model suffers from some tension caused by the correlation 
between $\Delta M_{s,d}$ and 
$|\varepsilon_K|$.

{\bf 1.} Indeed only small effects in $S_{\psi\phi}$ have been predicted, 
 which in 2008 was a problem that disappeared now.

{\bf 2.} The NP effects in $S_{\psi K_{S}}$ and $\Delta M_{d}/\Delta M_{s}$ 
turn out to be
very small
so that within this model 
these observables determine the coupling $V_{td}$, its phase $-\beta$ and its
magnitude $|V_{td}|$, without significant NP pollution. 
In particular we found 
$\gamma=63.5^{\circ}\pm 4.7^\circ$ and $|V_{ub}|\!=\! (3.5\pm 0.2)\cdot 10^{-3}$.
Thus in this model Scenario 1 for $\vub$ is favoured implying a value of
$|\varepsilon_K|$ that is visibly below the data if NP contributions are 
not included. 

{\bf 3.} Fortunately in this model $|\varepsilon_K|$ turns out to be uniquely enhanced over its SM value  and similarly to CMFV models also 
$\Delta M_{s,d}$ are enhanced in a correlated manner. Simply the plots in 
Fig.~\ref{fig:DeltaMvsepsK} apply here.
In 2008 it was possible to enhance  $\varepsilon_K$
up to a level of
$ 15\%$  basically removing  the corresponding anomaly but no definite 
statements could be made about $\Delta M_{s,d}$ due to large hadronic 
uncertainties present at that time.
In 2012, as 
$\Delta M_{d}/\Delta M_{s}$ is SM like in this model, it looks like the model is in a 
good shape at least from this point of view.However the first look 
at  $\varepsilon_K$ and the values of  $\Delta M_{d,s}$ indicates that 
not everything is optimal in this model. Indeed
taking 
the squarks of first two generations to be degenerate in mass and above 
$1~\tev$ and imposing 
the lower bound on the stop mass of $300~\gev$\footnote{We thank Wolfgang Altmannshofer for making this first look.}
shows that in 2012    $\varepsilon_K$
can only be enhanced by at most $7\%$ which only softens the problem with 
  $\varepsilon_K$ in the SM. Moreover then also   
$\Delta M_{d,s}$  increases 
automatically  by $10\%$ worsening the agreement with the data for these 
observables relative to the SM. 
Simply the correlation between $\Delta M_{d,s}$  and 
 $\varepsilon_K$ being in this model CMFV-like is not supported 
by the data. Still when
hadronic 
uncertainties are taken into account one cannot claim yet that this 
model fails to achieve consistency with the measured 
 $\Delta F=2$ observables.

Finally we notice that this model can be easily distinguished from
$SSU(5)_{\rm RN}$ on the basis of the value of $\gamma$ alone.

\subsection{The Minimal Effective Model with Right-handed Currents: RHMFV}

The recent phenomenological interest in making another look  
at the right-handed 
 currents in general, and not necessarily in the context of a given left-right 
 symmetric model, originated in tensions between inclusive and exclusive
 determinations of the elements of the CKM matrix $|V_{ub}|$ and  $|V_{cb}|$. 
 It could be that these tensions are due to the underestimate of theoretical 
 and/or experimental uncertainties. Yet, it is a fact, as pointed out
 and analyzed in particular in \cite{Crivellin:2009sd,Chen:2008se,Feger:2010qc,Crivellin:2011ba}, 
 that the presence of right-handed  currents could either remove 
 or significantly weaken some of these tensions, especially in the 
 case of $|V_{ub}|$. 

Assuming that RH currents provide the solution to
 the problem at hand, there is an important question whether the strength
 of RH currents required for this purpose is consistent with
 other observables and whether it implies new effects somewhere else that
 could be used to test this idea more globally.

This question has been addressed in \cite{Buras:2010pz}.
The starting point of this analysis is the assumption that 
the SM is the low-energy limit of a more
fundamental theory and consequently an effective theory is a useful 
approach to analyze the implications of RH currents.
In \cite{Buras:2010pz}
 the central role is played by  a left-right symmetric flavour group
 $SU(3)_L \times SU(3)_R$, commuting with 
an underlying $SU(2)_L \times SU(2)_R \times U(1)_{B-L}$ global symmetry and
broken only by two Yukawa couplings. The model contains a new 
unitary matrix $\Vt$  controlling flavour-mixing in the RH sector 
and can be considered as the minimally flavour violating generalization 
to the RH sector. Thus bearing in mind that this model contains non-MFV
interactions from the point of view of the standard MFV hypothesis that
includes only LH charged currents it can be called  RHMFV. 

It should be stressed from the start that similarly to 
 ${\rm 2HDM_{\overline{MFV}}}$
it is the high inclusive value 
of $\vub$ that is selected by the model as the true value of this element 
providing simultaneously the explanation of the smaller $\vub$ 
found in SM analysis of exclusive decays. The latter explanation is 
not offered in  ${\rm 2HDM_{\overline{MFV}}}$  but in both models the true large 
value of $\vub$
implies automatically a value 
of $\sin 2\beta$ above $0.80$ and therefore significantly larger 
than the measured value of $S_{\psi K_S}$. The question then arises 
how RHMFV solves this problem and what are its implications for 
$S_{\psi\phi}$ and $B_{s,d}\to \mu^+\mu^-$.

Now whereas the phenomenology of ${\rm 2HDM_{\overline{MFV}}}$ is governed by 
flavour blind phases in Yukawas and Higgs potential, this role in RHMFV 
is taken by the 
new mixing matrix $\Vt$  that can be parametrized in terms 
of 3 real mixing angles
and 6 complex phases. 
A detailed phenomenology of this matrix 
taking all tree level constraints into account and 
solving the $\vub$ problem in this manner allowed us to identify few 
favourite shapes for this matrix. Subsequently a detailed FCNC analysis 
has been performed \cite{Buras:2010pz}.

For our discussion of 2012 only the information about
 mixing structures relevant  to the three down-type $\Delta F=2$ 
and FCNC amplitudes in the RH sector is important.
Denoting by $\tc_{ij}$ and $\ts_{ij}$ the parameters of the RH matrix 
one finds  
that 
the $\tc_{12}$ and $\ts_{12}$ dependencies in the three systems considered are
non-universal with the observables in the 
$K$ mixing, $B_d$ mixing and $B_s$ mixing dominated by 
 $\tc_{12}\ts_{12}$,  $\tc_{12}$ and $\ts_{12}$, respectively. 

We should emphasize, probably for the first time, that this pattern allows 
to cope with the present data in a different manner than  
${\rm 2HDM_{\overline{MFV}}}$ does.

However, as 
both $\Delta S=2$ and $B_d$ mixing are strongly 
constrained, and the data from CDF and D0 gave some hints 
for sizable NP contributions to the CP violation in the $B_s$ mixing, 
it was  natural to assume in 2010 that $\tc_{12} \ll 1$. 
The phenomenological analysis was  then rather constrained but 
evidently with $\tc_{12} \ll 1$
the problem with the high value of $S_{\psi K_S}$ 
could not be removed. This forced the
authors of \cite{Buras:2010pz} to the following statement that we 
repeat here verbally:

{\it Thus our analysis casts a shadow on the explanation of the $|V_{ub}|$-problem 
with the help of RH currents alone unless the $S_{\psi\phi}$ anomaly goes 
away and $\tc_{12}$ can be large solving the problem with $S_{\psi K_S}$ 
naturally.}

 As of 2012 the RHMFV model of \cite{Buras:2010pz} is no longer under the shadow 
of a large value of $S_{\psi\phi}$. With its value given in (\ref{LHCb1}) 
the structure of the RH matrix changes and a large $\tc_{12}$ can be chosen 
 bringing  $S_{\psi K_S}$ down to its experimental value and introducing
only a small modification in $S_{\psi\phi}$.  It will be interesting to 
see whether RHMFV model works in detail when the LHCb data on $S_{\psi\phi}$ and 
other observables improve. However this  will require a new
numerical analysis. 

As far as the decays $B_{s,d}\to\mu^+\mu^-$ are concerned, 
already in 2010
the constraint from $B\to X_s \ell^+\ell^-$ precluded 
$\mathcal{B}(B_{s}\to \mu^+\mu^-)$ to be above  
$1\cdot 10^{-8}$.
Moreover NP effects in $B_{d} \to \ell^+\ell^-$ have been  found generally 
to be
smaller than in $B_{s} \to \ell^+\ell^-$. With the new structure of the RH 
matrix the opposite is true  and the NP effects in 
$B_{d} \to \ell^+\ell^-$ can now be much larger than in $B_{s} \to \ell^+\ell^-$
in accordance with 
the room left for NP in the LHCb data. 

Finally, let us mention that in this model, similar to ${\rm 2HDM_{\overline{MFV}}}$, the large value of $\vub$ softens significantly the problem with 
$B^+\to\tau^+\nu_\tau$.

There are other interesting consequences of this NP scenario that can be 
found in \cite{Buras:2010pz} even if some of them will be modified due 
to changes in the structure of the RH matrix. But let us stop here. It looks 
like the sun is again shining for RHMFV but it is not guaranteed that 
this will remain after new experimental informations will be available.

 \subsection{Left-Right Symmetric Models (LRM)}
The question then arises whether similar results can be obtained 
in a concrete BSM model with RH currents 
like the left-right  symmetric model (LRM) based 
on the weak gauge group $SU(2)_L\times SU(2)_R\times U(1)_{B-L}$
\cite{Pati:1974yy,Mohapatra:1974gc,Mohapatra:1974hk,Senjanovic:1975rk,Senjanovic:1978ev}. This question has been addressed in \cite{Blanke:2011ry} where
a complete study of $\Delta S=2$ 
and $\Delta B=2$ processes in a LRM 
including in particular $\varepsilon_K$, $\Delta M_{s,d}$
and the mixing induced CP asymmetries $S_{\psi K_S}$ and  $S_{\psi \phi}$ 
has been performed. Compared to the SM these observables 
are affected in this model by tree level
contributions from heavy neutral Higgs particles ($H^0$) as well as new box diagrams with $W_R$ gauge boson and charged Higgs 
$(H^\pm)$  exchanges. We also analysed the 
$B\to X_{s,d}\gamma$ decays that receive important new contributions from the 
$W_L-W_R$ mixing and 
$H^\pm$  exchanges. Compared to the previous literature the novel feature of our analysis 
was the  search for correlations between various observables that could  
help us to distinguish this model from other extensions of the 
SM and to obtain an insight into the structure of the mixing 
matrix $V^{\rm R}$ that governs right-handed currents. Moreover, we performed the 
full phenomenology including both gauge boson and Higgs boson contributions. We found that even for $M_{H^0}\approx M_{H^\pm}\sim\ord(20)\tev$, the tree level $H^0$ 
contributions to $\Delta F=2$ observables are by far dominant and the 
$H^\pm$ contributions to $B\to X_q\gamma$ can be very important, even dominant for certain parameters of the model.
While in a large fraction of the parameter space this  model has to struggle with 
the experimental constraint from $\varepsilon_K$,
we demonstrated that there exist regions in parameter space 
which satisfy all existing $\Delta F=2$, 
$B\to X_{s,d}\gamma$, tree level decays and electroweak precision constraints for scales 
$M_{W_R}\simeq 2-3\tev$ in the reach of the LHC.  We also showed that the 
$S_{\psi K_S}-\varepsilon_K$ tension present in the SM can be removed 
in the LRM. Simultaneously $\mathcal{B}(B\to X_s\gamma)$ can be brought closer 
to the data.
However, we pointed out that with the increased lower bound 
on $M_{W_R}$, the LRM cannot help in explaining the difference between 
the inclusive and exclusive determinations of $\vub$, when all constraints are 
taken into account, unless allowing for large fine-tuning.

 The present impact of a decreased value of $S_{\psi\phi}$ on this model 
can be significant in certain cases. In particular the allowed shape of 
the matrix $V^R$ is modified\footnote{Tillmann Heidsieck private communication.}. 

Also simple 
scenarios for this  matrix  considered in Section~7 of   
\cite{Blanke:2011ry} will be affected. However, to assess these changes 
an analysis of rare $K$ and $B$ decays in this model constraining better 
the free parameters of this model should be performed. The other 
general findings of this paper are still valid.

\subsection{A Randall-Sundrum Model with Custodial Protection}
Models with a warped extra dimension first proposed
by Randall and Sundrum (RS)  \cite{Randall:1999ee} provide a geometrical
explanation of the
hierarchy  between the Planck scale and the EW scale. Moreover, when the SM
fields, except for the Higgs field, are
allowed to propagate in the bulk 
\cite{Gherghetta:2000qt,Chang:1999nh,Grossman:1999ra}, 
these models naturally generate the
hierarchies in the fermion masses and mixing angles 
\cite{Grossman:1999ra,Gherghetta:2000qt} through different localisations 
of the fermions in the bulk. Yet, this way of explaining the hierarchies in masses 
and mixings necessarily   
implies FCNC transitions
at the tree level 
\cite{Burdman:2003nt,Huber:2003tu,Agashe:2004cp,Csaki:2008zd}.
Most problematic is the parameter 
$\varepsilon_K$ which receives tree level KK gluon contributions
 and some fine-tuning
of parameters in the flavour sector is necessary in order to achieve 
consistency with the data for KK scales in the reach of the LHC 
\cite{Csaki:2008zd,Blanke:2008zb}.

Once this fine-tuning is made,
the RS-GIM mechanism 
\cite{Huber:2003tu,Agashe:2004cp}, combined with an additional custodial
protection of  flavour violating $Z$ couplings 
\cite{Blanke:2008zb,Blanke:2008yr,Buras:2009ka}, 
 allows yet to achieve 
the agreement with
existing data for other observables 
without an additional fine tuning of parameters\footnote{See however comments 
in the final part of this subsection.}.
New theoretical ideas addressing the issue of large FCNC transitions in the
RS framework and proposing new protection mechanisms occasionally leading
to MFV can be found in 
\cite{Csaki:2008eh,Cacciapaglia:2007fw,Cheung:2007bu,Santiago:2008vq,Csaki:2009bb,Csaki:2009wc}.

In order to avoid problems
with electroweak precision tests (EWPT) and FCNC processes, the gauge group 
is
generally larger than the SM gauge group \cite{Agashe:2003zs,Csaki:2003zu,Agashe:2006at}:
\be
G_\text{RSc}=SU(3)_c\times SU(2)_L\times SU(2)_R\times U(1)_X
\ee
 and similarly to the LHT model
new heavy gauge bosons are present. The increased symmetry provides 
a custodial protection.  

The lightest new gauge bosons in this so-called RSc framework 
are the KK--gluons, the KK-photon and the 
electroweak KK gauge bosons $W^\pm_H$, $W^{\prime\pm}$, $Z_H$ and $Z^\prime$,
all with masses $M_{KK}$ around $2-3\tev$ as required by the consistency 
with the EWPT \cite{Agashe:2003zs,Csaki:2003zu,Agashe:2006at}.
 The fermion sector is 
enriched through heavy KK-fermions (some of them with exotic electric charges)
 that could in principle be discovered at 
the LHC. The fermion content
of this model is explicitly given in \cite{Albrecht:2009xr}, where also 
 a complete set of 
Feynman rules has been worked out. Detailed analyses of electroweak precision
tests and of the parameter $\varepsilon_K$ in a RS model without and with 
custodial 
protection can also be found in \cite{Casagrande:2008hr,Bauer:2008xb}. These authors 
analyzed also rare and non-leptonic decays in \cite{Bauer:2009cf}.
Possible flavour protections
in warped Higgsless models have been presented in  \cite{Csaki:2009bb}. 

We will now summarize the impact of LHCb data on the results presented 
in \cite{Blanke:2008zb,Blanke:2008yr}:

{\bf 1.} The CP asymmetry $S_{\psi\phi}$ was found in RSc in 2008 to 
reach values as high as  0.8. Such values are clearly excluded at 
present but RSc can have also SM-like values for this asymmetry and 
if necessary the asymmetry can be negative.

{\bf 2.} The smallness of  $S_{\psi\phi}$ are good news for rare $K$ 
decays as in RSc simultaneous large NP effects in $S_{\psi\phi}$
          and $K\to\pi\nu\bar\nu$ channels are very unlikely and this 
          feature is even more pronounced than in the LHT model.
Thus as stated by the authors of \cite{Blanke:2008zb,Blanke:2008yr} on 
many occasions:
SM-like value of $S_{\psi\phi}$
 would open the road to large enhancements of these branching ratios that
 could be tested by ${\rm K^0}$TO at J-Parc, NA62 at CERN and ORCA at Fermilab. It 
looks like LHCb opened this road this year.
        
{\bf 3.} Indeed in the absence of a large  $S_{\psi\phi}$ the branching ratios 
          for $\kpn$, $\klpn$, $K_L\to\pi^0\ell^+\ell^-$
         can be enhanced relative to the SM expectations up to factors of
         1.6, 2.5 and 1.4, respectively, when only moderate fine tuning 
in $\varepsilon_K$ is required. Otherwise the enhancements can be larger.
          $\mathcal{B}(\kpn)$ and $\mathcal{B}(\klpn)$ can
         be simultaneously enhanced but this is not necessary as the
         correlation between these two branching ratios is not evident
         in this model. On the other hand $\mathcal{B}(\klpn)$ and 
         $\mathcal{B}(K_L\to\pi^0\ell^+\ell^-)$ ($\ell=e,\mu)$ are strongly correlated and the
         enhancement of one of these three branching ratios implies the
         enhancement of the remaining two.

{\bf 4.} The branching ratios for $B_{s,d}\to \mu^+\mu^-$ and 
             $B\to X_{s,d}\nu\bar\nu$ remain SM-like: the maximal enhancements
            of these branching ratios amount to $15\%$. This is clearly 
            consistent with the present LHCb data but the situation may 
            change in the future.

{\bf 5.} The CMFV relations (\ref{dmsdmd}),  (\ref{bxnn}), 
(\ref{bmumu}) and  (\ref{R1})    between various observables 
         can be strongly
         violated.  

 Next, let us just mention  that large NP contributions in the RS framework
 that require some tunings of parameters
in order to be in agreement with the experimental data have been found in 
$\mathcal{B}(B\rightarrow X_s\gamma)$ \cite{Agashe:2008uz}, $\mathcal{B}(\mu\rightarrow
 e\gamma)$ \cite{Agashe:2006iy,Davidson:2007si,Agashe:2009tu} and EDMs
 \cite{Agashe:2004cp,Iltan:2007sc}, that are all dominated by dipole operators. Also the new 
contributions to $\varepsilon^{\prime}/{\varepsilon}$ can
be large \cite{Gedalia:2009ws}. Moreover it appears that the fine tunings 
in this ratio are not necessarily consistent with the ones required in the 
case of $\varepsilon_K$.

Finally, we would like to mention a very recent study of $B\to X_s\gamma$, 
$B\to K^*\mu^+\mu^-$, $B\to K^*\gamma$ and of the related observables
in the RSc model \cite{Blanke:2012tv}. As these processes were not 
the main stars of the present review we refer to a very systematic  summary section of this paper for details.

Now  many of the ideas and
concepts that characterize most of the physics discussed in the context 
of RS scenario  
do not rely on the assumption
of additional dimensions and
as indicated by AdS/CFT correspondence
 we can regard RS models
as a mere computational tool for certain 
strongly coupled theories. Therefore in spite of some tensions in this NP 
scenario, the techniques developed in the last decade will 
certainly play an important role in the phenomenology in particular if 
Higgs will not be found to be an elementary particle and a new strong dynamics 
will show up at the LHC.

\subsection{Gauged Flavour Models}

In \cite{Grinstein:2010ve,Feldmann:2010yp,Guadagnoli:2011id} 
a MFV-like ansatz has been  implemented in the context of maximal gauge flavour (MGF) symmetries: in the limit
of vanishing Yukawa interactions these gauge symmetries are the largest non-Abelian ones allowed by the Lagrangian
of the model. The particle spectrum is enriched by
new heavy gauge bosons, carrying neither colour nor electric charges, and exotic fermions,
to cancel anomalies. Furthermore, the new exotic fermions give rise to the SM fermion
masses through a seesaw mechanism, in a way similar to how the light left-handed (LH) neutrinos
obtain masses by the heavy RH ones. Moreover, the MFV spurions are promoted to scalar fields -- called flavons --
invariant under the gauge group of the SM, but transforming as bi-fundamental representations of
the non-Abelian part of the flavour symmetry. Once the flavons develop suitable VEVs, the SM fermion
masses and mixings are correctly described. 

Even if this approach has some similarities to the usual MFV description, the presence of 
flavour-violating neutral gauge bosons and exotic fermions introduces modifications of the SM couplings and
tends to lead to dangerous contributions to FCNC processes mediated by the new heavy particles.
Consequently, the MGF framework goes beyond the standard MFV and a full phenomenological analysis 
of this NP scenario is mandatory to judge whether it is consistent with all available data.

In \cite{Buras:2011wi} we have presented a detailed analysis
of $\Delta F=2$ observables 
and of $B\to X_s\gamma$ in the framework of a specific 
MGF model of 
Grinstein {\it et al.}  \cite{Grinstein:2010ve} 
including all relevant contributions, in particular tree-level heavy gauge boson exchanges.
The number of parameters in this model is much smaller than in
some of the extensions of the SM discussed above and therefore it is not obvious
that the present tensions on the flavour data can be removed or at least softened. Our findings are as follows:

{\bf 1.}
We find that large corrections to the CP observables in the meson oscillations, $\varepsilon_K$,
$S_{\psi K_s}$ and $S_{\psi\phi}$, are allowed. However, requiring 
$\varepsilon_K$ to be in agreement with experiment
only small deviations from the SM values of $S_{\psi K_s}$ and $S_{\psi\phi}$ 
are allowed. While at the time of our analysis this appeared as a possible 
problem as far as  $S_{\psi\phi}$ was concerned, this result is now fully consistent with present LHCb data.

{\bf 2.}
Consequently we find that this model selects the scenario with 
exclusive (small) value of $|V_{ub}|$.  

{\bf 3.}
$|\varepsilon_K|$ is enhanced without modifying $S_{\psi K_S}$.

{\bf 4.}
 The values of $\Delta M_{d}$ and $\Delta M_{s}$ being strongly
correlated in this model with $\varepsilon_K$ turned  out to be 
enhanced. In our original paper, were 2011 lattice input has been used, they 
were much larger than the data for the central
values of input parameters: $\Delta M_{d}\approx0.75~\text{ps}^{-1}$ and $\Delta M_{s}\approx27~\text{ps}^{-1}$. Meanwhile the
lattice values for the
relevant non-perturbative 
parameters have been modified so that this problem has been softened:
$\Delta M_{d}\approx0.69~\text{ps}^{-1}$ and $\Delta M_{s}\approx 23.9~\text{ps}^{-1}$ as in CMFV 
(see (\ref{BESTCMFV}).
Therefore after the inclusion 
of theoretical and parametric uncertainties  these central values are within 
$3\sigma$ from the data. Further decrease of 
non-perturbative uncertainties is necessary to fully assess whether 
this model fails to describe the data on $\Delta F=2$ processes properly.

{\bf 5.}
Also problematic for this model at present appears to be the branching ratio 
$\mathcal{B}(B^+\to\tau^+\nu_\tau)$ for which the model does not provide 
any improvement with respect to the SM.

In summary the new LHCb data and new lattice input provided a 
 relief for MGF as far as $S_{\psi\phi}$ is concerned but a sactisfactory 
simultaneous description of $\Delta M_{s,d}$ and $|\varepsilon_K|$ has 
not been yet achieved in this model.
 Also the experimental value of 
$\mathcal{B}(B^+\to\tau^+\nu_\tau)$ remains still a big problem due to 
the small $\vub$ selected by this model. Yet, $\mathcal{B}(B^+\to\tau^+\nu_\tau)$ could turn out to be smaller one day.

\begin{figure}[!tb]
\centerline{\includegraphics[width=0.65\textwidth]{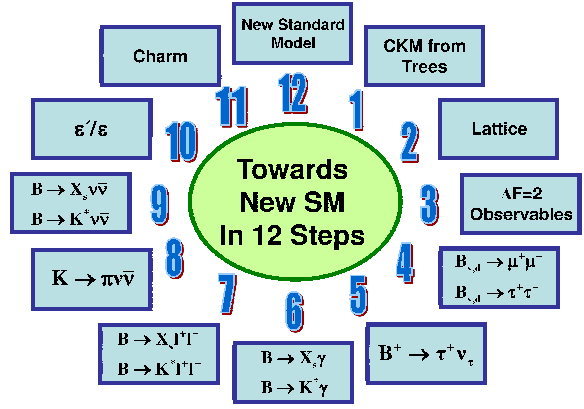}}
\caption{Towards the New Standard Model in 12 Steps.}\label{Fig:1}
\end{figure}

\section{Observations, Messages and a Shopping List}\label{sec:4}
Our BSM story is approaching the end. We have seen that in certain cases 
the recent LHCb data on $S_{\psi\phi}$ and $B_{s,d}\to\mu^+\mu^-$ had a profound 
impact on some extensions of the SM. Also most recent update on lattice input 
modified predictions for $\Delta M_{s,d}$ and $B^+\to\tau^+\nu_\tau$  not only 
within several BSM scenarios but also within the SM. It is to be expected 
that in the coming years the new data from LHCb combined with direct searches for NP at the LHC and further improvements coming
from lattice calculations will 
have an important impact on the landscape of BSM scenarios reviewed here, 
reducing it to a few oases. We should then hope that the measurements 
performed during the second LHC phase, by upgraded LHCb and in 
particular SuperKEKB, SuperB in Rome and Kaon physics dedicated experiments 
NA62, ${\rm K^0TO}$ and ORCA will select one oasis represented by a new 
SM (NSM). For recent reviews see \cite{Ciuchini:2011ca,Komatsubara:2012pn}.

The route to the NSM will not be easy and will involve within the quark 
flavour physics at least {\it twelve} steps depicted in Fig.~\ref{Fig:1}. 
This route will be complemented by another one involving the lepton flavour 
violation and if we are lucky the two routes will meet in the NSM oasis. 
But the description of these routes is another story.

For the coming years we are less ambitious. Nevertheless our shopping 
list includes some steps that in 2020 will surely be classified as 
mile stones in quark flavour physics.

Our shopping list has been constructed on the basis of what we have seen on previous
pages and we summarize here our observations and related messages:
\begin{itemize}
\item
 First of all we have identified a number of models that select either 
small or large value of $\vub$: they simply have only a chance to describe the $\Delta F=2$ 
observables properly for such values. These are models in which NP can contribute 
significantly to either $\varepsilon_K$ or $S_{\psi K_S}$ but not to both of 
them. 
The small value of $\vub$ is chosen by CMFV, MFV without FBPhs, FBMSSM and 
 MGF models. Large value of $\vub$ is selected by 
${\rm 2HDM_{\overline{MFV}}}$ and RHMFV. Thus already clarification what is 
the true value of $\vub$ will tell us which of these two classes of models 
should be favoured.
\item
There are models like $SSU(5)_{\rm RN}$ in which there are no significant 
contributions to  $\varepsilon_K$ or $S_{\psi K_S}$. Such models can only 
provide satisfactory description of the $\Delta F=2$ data through an increased value of 
$\gamma$, typically above $80^\circ$. Therefore future tree level 
measurements of $\gamma$ will tell us whether such models offer good 
description of $\Delta F=2$ data or not. As we remarked in the corresponding 
section, from the present perspective we do not expect this to be the case.
\item
We have seen that in models with CMFV and MFV without FBPhs,  $S_{\psi\phi}$ 
had the SM value. On the other hand in ${\rm 2HDM_{\overline{MFV}}}$  it 
could have a different value but only a {\it positive} one and larger than the 
SM one. Models with new sources of flavour violation like LHT, SM4, 
SF models with RH-currents and the CMM model could generate $S_{\psi\phi}$ with both signs. 
Finding $S_{\psi\phi}$ to be {\it negative} would clearly indicate 
new sources of flavour and CP violation at work.
\item
The ratio $\Delta M_s/\Delta M_d$ within SM, CMFV, ${\rm 2HDM_{\overline{MFV}}}$ 
and MGF models is roughly the same and consistent with 
experiment. 
On the other hand while in SM the values of $\Delta M_s$ and  $\Delta M_d$ 
are only slightly above the data, the desire to lift up 
$|\varepsilon_K|$ over its SM value shifts automatically 
 $\Delta M_s$ and  $\Delta M_d$ in CMFV and MGF models so that we have 
to conclude that
in these models it is not possible
to obtain a good fit simultaneously to $\Delta M_{s,d}$ 
and $\varepsilon_K$. This problem does not exist in  ${\rm 2HDM_{\overline{MFV}}}$ because there is no correlation between  $\Delta M_{s,d}$ 
and $\varepsilon_K$ in this model.
\item
As seen in Fig.~\ref{fig:BsmumubsBdmumu}, $B_{s,d}\to\mu^+\mu^-$ can 
distinguish between various scenarios, in particular when considered 
simultaneously. In  SM4 and SF models with LH-currents 
$\mathcal{B}(B_d\to\mu^+\mu^-)$ can still be enhanced by one order of 
magnitude with respect to the SM value. We have also found that in SM4 
the suppression of $\mathcal{B}(B_s\to\mu^+\mu^-)$ relative to the SM is 
rather likely, while in the LHT model only enhancement is possible and in
the CMM model both decays  $B_{s,d}\to\mu^+\mu^-$ stay SM-like. 
Evidently already the enhancements or suppressions of 
$\mathcal{B}(B_{s,d}\to\mu^+\mu^-)$ with respect to SM values will select 
favourite scenarios of NP.
In this context the CMFV relations 
(\ref{dmsdmd})-(\ref{R1}), (\ref{NonDirect}) and other 
relations discussed in \cite{Buras:2003jf} can be regarded as {\it 
standard candels of flavour physics}  and the deviations from them may help 
in identifying the correct NP scenario.
\item
Concerning $B^+\to\tau^+\nu_\tau$ we do not yet take the discrepancy 
with the SM prediction as seriously as some authors do. It should be 
kept in mind that these are the first measurements of this decay and 
it could well be that its branching ratio is much closer to the SM value. 
In this context precise determinations of $\vub$ and $F_{B^+}$ are 
very important.
\item
Our discussion of $\kpn$, $\klpn$ and $B\to K^*\ell^+\ell^-$ was marginal 
in view of space imitations but these decays together with $b\to s\nu\bar\nu$ 
transitions are among super stars of this decade.
\end{itemize}

Having all this in mind our
 shopping list for the coming years looks as follows:
\begin{itemize}
\item 
 Clarification of the discrepancy between inclusive and exclusive 
values of $\vub$.
\item
 Precise measurement of the angle $\gamma$ in tree-level decays.
\item
Improved lattice input for $\Delta F=2$ observables.
\item
Precise measurement of $S_{\psi\phi}$, in particular the determination 
of its sign.
\item  
Improved calculations of $\Delta M_s$ and $\Delta M_d$ in order to 
test CMFV relations (\ref{dmsdmd}) and (\ref{R1}).
\item
Improved measurements of $\mathcal{B}(B_{s,d}\to\mu^+\mu^-)$ with 
the goal to test the MFV relation (\ref{bmumu}).
\item
Improved measurement of $\mathcal{B}(B^+\to\tau^+\nu_\tau)$ 
combined with improved $\vub$ and weak decays constant $F_{B^+}$.
\item
Measurements of the branching ratio $\mathcal{B}(\kpn)$ by  NA62 at CERN 
and ORCA at Fermilab and of  $\mathcal{B}(\klpn)$ by ${\rm K^0}$TO.
\item
Precise study of angular observables in $B\to K^*\ell^+\ell^-$.
\end{itemize}

We could continue like this around the clock in Fig.~\ref{Fig:1} but let 
us stop here. Certainly the coming years will be exciting for flavour 
physics.

{\bf Acknowledgements}\\
The first author would like to thank the organizers for the possibility 
of giving this talk at such well organized and interesting conference.
The discussions with Wolfgang Altmannshofer, Monika Blanke, Katrin Gemmler, Tillmann Heidsieck, Luca Merlo,
Minoru Nagai, Paride Paradisi,
Stefan Recksiegel, Emmanuel Stamou and David Straub are highly appreciated. 
We would also like to thank Christine Davies, Andreas Kronfeld, Jack Laiho, Enrico Lunghi and Ruth Van de Water for detailed information about the lattice 
input. 
This research was done in the context of the ERC Advanced Grant project ``FLAVOUR''(267104) and was partially supported by the DFG cluster
of excellence 
``Origin and Structure of the Universe'' and  by the German ``Bundesministerium f\"ur Bildung und Forschung'' under contract
05H09WOE.



\end{document}